\documentclass[reprint,twocolumn,footinbib,longbibliography,superscriptaddress]{revtex4-2}
\usepackage{amsmath}
\usepackage{amsfonts}
\usepackage{amssymb}
\usepackage{lmodern}
\usepackage{times}
\usepackage[pdftex]{graphicx}
\usepackage{bm}
\usepackage{amsthm}
\usepackage{mathtools}
\usepackage{physics}
\usepackage{mathrsfs}
\usepackage{hyperref}
\usepackage{url}
\usepackage{color}

\newcommand{\Income}{{{\bf I}}}
\newcommand{\Metabo}{{\mathcal{M}}}
\newcommand{\Reac}{{\mathcal{R}}}
\newcommand{\vflux}{{\bf v}}
\newcommand{\Exchange}{{\mathcal{E}}}
\newcommand{\BM}{\mathrm{gr}}
\newcommand{\Const}{{\mathcal{C}}}
\newcommand{\Nint}{\mathsf{N}^{\rm int}}
\newcommand{\Nex}{\mathsf{N}^{\rm ex}}
\newcommand{\yec}{{{\bf y}_{\mathcal{E}\cup\mathcal{C}}}}

\begin{document}
\preprint{APS/123-QED}
\title{
    Global Constraint Principle for Microbial Growth Law
}
\author{Jumpei F. Yamagishi}
 \email{jumpei.yamagishi@riken.jp}
 \affiliation{Center for Biosystems Dynamics Research, RIKEN, 6-2-3 Furuedai, Suita 565-0874, Japan}
 \affiliation{Universal Biology Institute, The University of Tokyo, 7-3-1 Hongo, Tokyo 113-0033, Japan}
\author{Tetsuhiro S. Hatakeyama}
 \email{hatakeyama@elsi.jp}
 \affiliation{Earth-Life Science Institute, {Institute of Science Tokyo}, Tokyo 152-8550, Japan}

\begin{abstract}
Monod's law is a widely accepted phenomenology for bacterial growth. Since it has the same functional form as the Michaelis--Menten equation for enzyme kinetics, cell growth is often considered to be locally constrained by a single reaction. In contrast, this Letter shows that a global constraint {principle} of resource allocation to metabolic processes can well describe the nature of cell growth. This concept is a generalization of Liebig's law, a growth law for higher organisms, and explains the dependence of microbial growth on the availability of multiple nutrients, in contrast to Monod's law. 
\end{abstract}

\maketitle

To understand complex living systems, universal phenomenological laws independent of specific species or molecules are of paramount importance~\cite{kaneko2006life}. 
In particular, the dependence of cellular growth rates on environmental conditions (and cellular physiological states) has been an important problem in biology for over a century~\cite{monod1949growth, whitson1912soils,scott2011bacterial,kovarova1998growth}. 
The classical phenomenology, originally proposed by Monod in the 1940s~\cite{monod1949growth}, states that microbial growth kinetics, namely the dependence of the specific growth rate $\mu$ on substrate availability, generally follows the Monod equation: 
\begin{eqnarray} \label{eq:Monod}
    \mu([S]) = \mu_{\max}\frac{[S] }{K_S+[S]},
\end{eqnarray}
where $\mu_{\max}$ is the maximum growth rate and $[S]$ is the environmental concentration of the growth-limiting substrate $S$ with its half-saturation concentration $K_S$. 

\begin{figure}[t]
    \centering 
    \includegraphics[width=0.99\linewidth, clip]{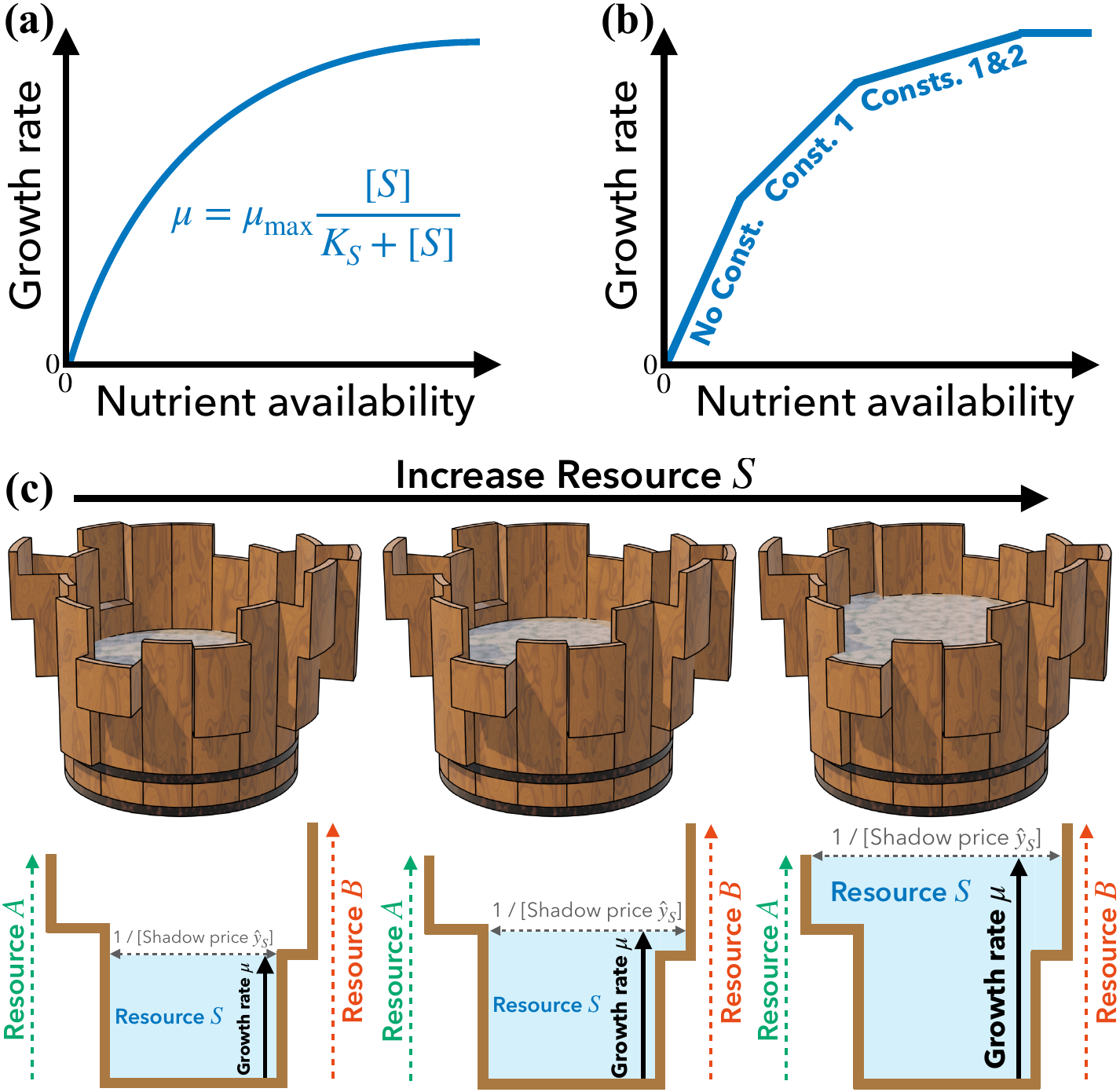}
    \caption{
    Microbial growth kinetics. (a) Monod's microbial growth model. 
    (b) Global constraint {principle} for the microbial growth kinetics curve. 
    (c) Terraced Liebig's barrel (top) and its sectional view for the case with a focal substrate $S$ and the other resources $A$ and $B$ (bottom). 
    As $S$ is poured into the barrel, the surface height corresponds to the growth rate $\mu$, {where} the height of each stave represents the availability of each resource other than $S$. {The water surface area in the barrel is therefore $\partial I_S/\partial \mu$, which coincides with the inverse of the shadow price $\hat{y}_S\,{=}\,\partial \mu/\partial I_S$ (i.e., the slope of the growth kinetics curve).} Each stave spreads out in a stepwise manner, reflecting the metabolic shifts to pathways with lower growth yield from $S$ due to additional global constraints. 
    }\label{fig:Schematics}
\end{figure}

Such behavior has often been attributed to a specific molecular biological process. Since the Monod equation~\eqref{eq:Monod} has the same functional form as the Michaelis--Menten equation, a single biochemical process is usually {assumed} to limit cell growth~\cite{reich1981energy,sher2024quantitative,liu2007overview,de2017mathematical} (see also Fig.~\ref{fig:Schematics}a). 
In previous hypotheses, this single ``black box'' could be the substrate transport into cells~\cite{merchuk1995monod, reich1981energy}, the net flux of respiration~\cite{jin2003new}, or the (coarse-grained) reaction that couples catabolism and anabolism~\cite{heijnen1995derivation}.

However, microbial growth is achieved collectively by the interplay of thousands of biochemical processes. There is considerable experimental evidence that cell growth is not limited by a single biochemical process alone~\cite{koch1997microbial,goelzer2011bacterial}: for example, experimental changes in the environmental conditions of nitrogen sources can alter the dependence of growth rate on the availability of carbon sources~\cite{bren2016glucose,tang2021finding}; and cells exhibit qualitatively different phenotypes with respect to the metabolism of nutrients for cell growth~\cite{palsson2015systems}. 

In addition, the accumulation of experimental data has shown that the experimentally-observed microbial growth kinetics are not accurately captured by the Monod equation~\eqref{eq:Monod}~\cite{reich1981energy,sommer1991comparison,kovarova1998growth,wang2022mathematical}. That is, the shape of actual microbial growth kinetics curves varies across species and environments and there is no longer quantitative evidence to support the exact functional form of Eq.~\eqref{eq:Monod}. Nevertheless, certain characteristics still appear to be universal: (i) growth rate is a monotonically increasing function of nutrient availability and (ii) growth rate exhibits concavity with respect to nutrient availability~\footnote{
    Empirically, there can be a finite minimum substrate concentration $[S]_{\min}$ to achieve cell growth due to the maintenance energy requirements~\cite{kovarova1998growth, goelzer2011bacterial}. 
    However, in classical arguments such as Monod's, they are effectively ignored for simplicity by considering the substrate concentration $[S]$ after subtracting $[S]_{\min}$.}, 
in other words, there are diminishing returns to getting more and more of a nutrient.

Therefore, a novel macroscopic framework is essential to elucidate the general mechanism underlying these fundamental characteristics (i-ii) of the microbial growth kinetics curve. Based on the optimality principle of evolved metabolic systems, we show that the cellular growth rate gradually plateaus with increasing nutrient availability. This is because as the availability of a specific nutrient increases, some other resources are gradually depleted and thus act as additional growth-limiting factors (Fig.~\ref{fig:Schematics}b). 
This framework can be seen as a generalization of Liebig's law of the minimum, a classical phenomenological theory of plant growth, which states that the availability of multiple resources collectively limits the growth of organisms~\cite{von1840organic,whitson1912soils,tang2021finding} (see also Fig.~\ref{fig:Schematics}c). It also provides new insights into the dependence of microbial growth rate on the environmental availability of multiple nutrient sources, which cannot be captured by Monod's classical phenomenology.

\textit{Global constraint {principle} of cell growth.---}
Cell growth requires the production of biomass through metabolism. We therefore focus on how the intracellular allocation of diverse resources to metabolic processes is globally constrained by stoichiometry and the law of matter conservation. 

Let us denote all the intracellular metabolites and metabolic reactions as $\Metabo$ and $\Reac$, respectively. Here, by decomposing each reversible reaction into two irreversible reactions (i.e., its forward and backward components), a non-negative vector $\vflux\,{:=}\,\{v_i\}_{i\in\Reac}$ represents the fluxes of all $|\Reac|$ reactions. 
Among $|\Metabo|$ metabolites, the set of exchangeable metabolites is denoted by $\Exchange\,({\subset}\,\Metabo)$. 

In the context of systems biology, the metabolic systems of cells are assumed to maximize their growth rate, or the biomass synthesis rate, as a natural consequence of evolution~\cite{palsson2015systems,lewis2012constraining,ibarra2002escherichia,klipp2016systems}. Since we are considering continuously growing microbes, stationarity of intracellular concentrations of non-exchangeable metabolites $m\,({\in}\,\Metabo\backslash\Exchange)$ is also assumed. 
Then, the metabolic regulation of reaction fluxes $\vflux$ can generally be formulated as a linear programming (LP) problem, as known as constraint-based modeling (CBM) of metabolism~\cite{palsson2015systems,warren2007duality,yamagishi2023linear}: 
\begin{eqnarray} \label{eq:CBM}
    \underset{\vflux\geq{\bf 0}}{{\rm maximize}} 
    \;\;v_\BM \;\; \mathrm{s.t.} 
    &&\sum_{i\in\Reac}\mathsf{S}_{m i}v_i=0\;\;(m\in\Metabo\backslash\Exchange), \label{eq:Balanced} \\
    &&\sum_{i\in\Reac}\mathsf{S}_{m i}v_i + I_m \geq 0\;\;(m\in\Exchange),  \label{eq:Leakage} \\
    &&\sum_{i\in\Reac}\mathsf{C}_{a i} v_i \leq I_a  \;\;(a\in\mathcal\Const), \label{eq:ResourceAllocation}
\end{eqnarray}
where $\BM\,({\in}\,\Reac)$ denotes the growth or biomass synthesis reaction. $\mathsf{S}_{m i}$ is the stoichiometric coefficient representing that metabolic reaction $i\,({\in}\,{\Reac})$ produces (consumes) $|\mathsf{S}_{m i}|$ molecules of metabolite $m\,({\in}\,{\Metabo})$ if $\mathsf{S}_{m i}\,{>}\,0$ (if $\mathsf{S}_{m i}\,{<}\,0$). $\sum_{i}\mathsf{S}_{m i}v_i$ is then the excess production of metabolite $m$. 
Thus, equalities~\eqref{eq:Balanced} represent that the production and degradation of internal metabolites must be balanced, and inequalities~\eqref{eq:Leakage} represent that exchangeable metabolites $m$ with maximal influx $I_m\,{>}\,0$ (efflux $I_m\,{<}\,0$) are taken up (degraded). 
{The} other constraints on the allocation of the non-nutrient resources ${m\,({\in}\,{\Const})}$ {are incorporated as inequalities~\eqref{eq:ResourceAllocation}: $I_m$ represents the total capacity of non-nutrient resource $m$, such as} the total amount of enzymes~\cite{Hwa_OM,mori2016constrained} {or} membrane surface {area}~\cite{memRealEstate2} {or the upper bound on the} Gibbs energy dissipation~\cite{Heinemann-Gibbs2019}. 
{Therefore, the LP problem~(\ref{eq:Balanced}-\ref{eq:ResourceAllocation}) includes standard flux balance analysis~\cite{palsson2015systems,klipp2016systems} and its variants~\cite{mori2016constrained,beg2007intracellular,vazquez2008impact,Heinemann-Gibbs2019,carlson2024cell} as well as resource balance analysis~\cite{goelzer2011cell} as special cases.}
Table~S1 collects all the notation in one place. 

The solution to the LP problem~(\ref{eq:Balanced}-\ref{eq:ResourceAllocation}) is denoted as $\hat{\vflux}(\Income)$; since the optimal objective function $\hat{v}_\BM(\Income)$ is the predicted value of the growth rate $\mu(\Income)$, we will hereafter refer to $\hat{v}_\BM(\Income)$ as $\mu(\Income)$. 
Numerical calculations of the CBM method successfully explain and predict experimental data, at least for model organisms such as \textit{E. coli} strains~\cite{palsson2015systems,lewis2012constraining,ibarra2002escherichia,shinfuku2009development}.

\textit{Monotonicity and concavity of growth kinetics curve.---}
The calculated growth rate satisfies fundamental characteristics (i-ii) of the microbial growth kinetics curve. Here we show that the dependence of the growth rate $\mu$ on the maximal influx $I_S$ of a given nutrient $S\,({\in}\,\Exchange)$ is generally a monotonically increasing and concave function. 
While such properties have sometimes been recognized empirically for specific metabolic models and parameters~\cite{zeng2020bridging,yamagishi2021microeconomics}, its generality can be proven mathematically as follows.  

First, the monotonic increase in the growth kinetics curve $\mu(I_S;\tilde{\Income})$, where $\tilde{\Income}\,{:=}\,\{I_a\}_{a\in\Exchange\cup\Const\backslash\{S\}}$ denotes the availability of the resources other than the nutrient $S$, is evident. 
{
    When the maximal nutrient influx increases from $I_S$ to $I_S\,{+}\,\Delta{I_S}$ ($\Delta{I_S}\,{>}\,0$), any $\vflux$ satisfying inequalities~\eqref{eq:Leakage} also satisfies $\sum_i \mathsf{S}_{Si}v_i\,{+}\,I_S\,{+}\,\Delta{I_S}\,{\geq}\,0$. 
    That is, any originally feasible solution $\vflux$ that satisfies  conditions (\ref{eq:Balanced}-\ref{eq:ResourceAllocation}) remains a feasible solution even after increasing nutrient availability, and thus increasing $I_S$ monotonically expands the feasible solution space. 
    The growth rate $\mu(I_S;\tilde{\Income})\,{:=}\,\hat{v}_\BM(\Income)$ is therefore a monotonically increasing function of $I_S$.} 

To prove the concavity of the growth kinetics curve $\mu(I_S;\tilde{\Income})$, it suffices to show that its slope, $ \partial \mu/\partial I_S $, monotonically decreases as $I_S$ increases. 
Remarkably, the slope coincides with the so-called shadow price $\hat{y}_S$ of nutrient $S$~\cite{reznik2013flux,varma1993stoichiometric,edwards2002characterizing}, where $\hat{{\bf y}}$ denotes the solution to the dual problem defined below and its element $\hat{y}_a$ quantifies the growth return to the additional consumption of resource $a\,({\in}\,\Exchange\,{\cup}\,\Const)$. In the following, we show that shadow price $\hat{y}_S(I_S;\tilde{\Income})$ diminishes monotonically with $I_S$.

\begin{figure*}[htb]
    \centering
    \begin{minipage}{0.7\textwidth} 
        \centering
        \includegraphics[width=\linewidth]{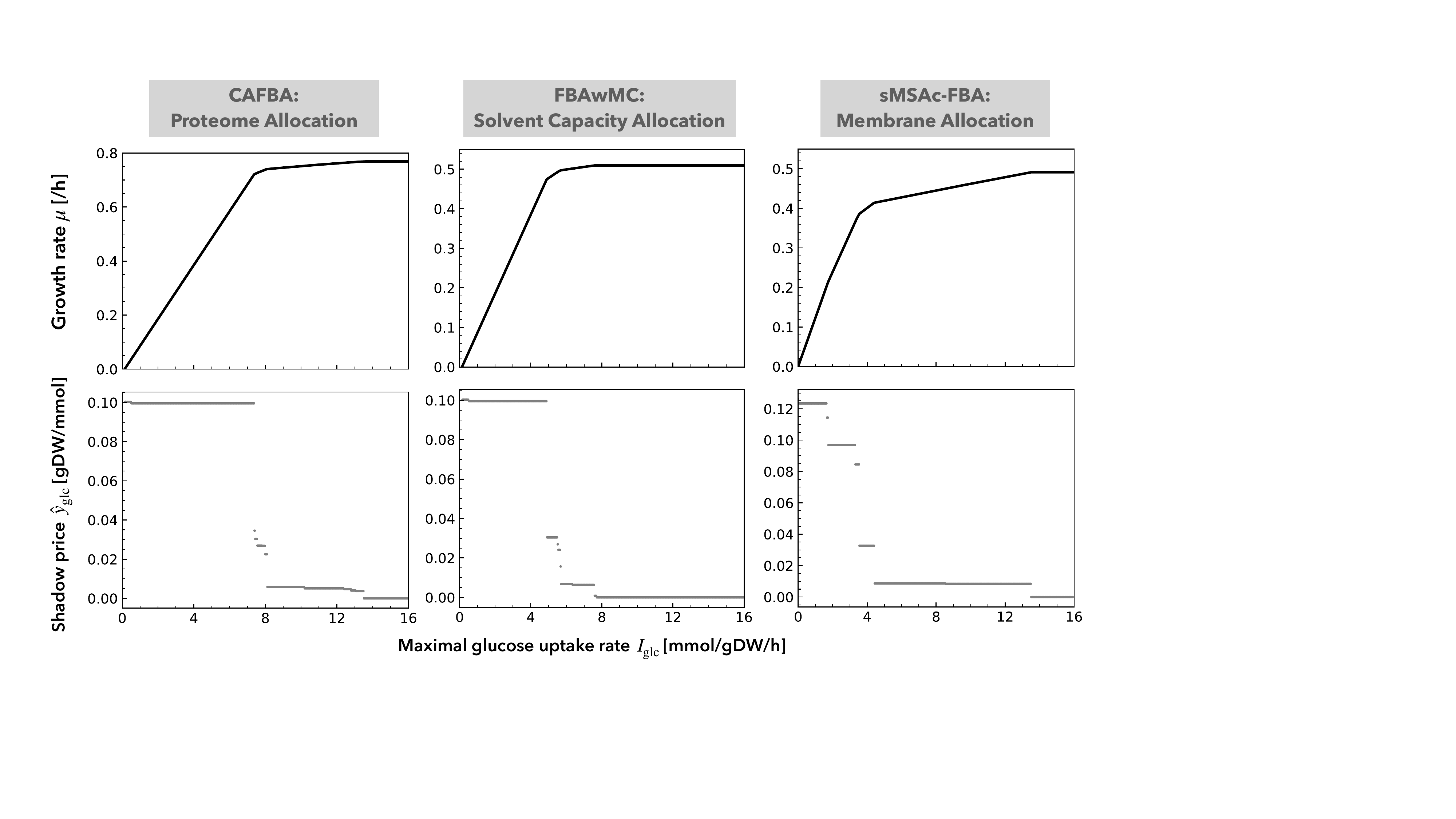} 
    \end{minipage}
    \hspace{7pt} 
    \begin{minipage}{0.275\textwidth} 
        \vspace{0pt}
        \caption{
            {
            Growth rate $\mu$ (top) and shadow price $\hat{y}_{\rm glc}$ of glucose (bottom) as a function of carbon source availability $I_\mathrm{glc}$. 
            Numerically calculations of various CBM methods with either constraint on the allocation of proteome (constrained allocation flux balance analysis, CAFBA~\cite{mori2016constrained}), cytoplasmic volume (FBA with molecular crowding, FBAwMC~\cite{beg2007intracellular,vazquez2008impact}), or membrane surface area (specific membrane surface area-constrained FBA, sMSAc-FBA~\cite{carlson2024cell}) were performed (see Appendix~\ref{sec:CBMs} for details). 
        }
        } \label{fig:VariousFBAs}
    \end{minipage}
\end{figure*}

The dual problem to the primal LP problem~(\ref{eq:Balanced}-\ref{eq:ResourceAllocation}) is given as~\cite{warren2007duality,vanderbei1998linear,reznik2013flux} (see also Appendix~\ref{sec:dual} for the derivation {and a simple example}):
\begin{eqnarray}
    \underset{{\bf y}\in\mathbb{R}^{\Metabo\cup\Const}}{{\rm minimize}}
    \sum_{a\in\Exchange\cup\Const} I_a y_a
    \;\; \mathrm{s.t.} &&
    \begin{pmatrix}
        -{\mathsf{S}}^\top & {\mathsf{C}}^\top 
    \end{pmatrix} {\bf y} \geq 
    {\bf 1}_\BM, \label{eq:dual}\\
    && y_a\geq 0\;\;(a\in\Exchange\cup\Const). \nonumber
\end{eqnarray}
Here ${\bf 1}_\BM$ denotes a $|\Reac|$-dimensional vector, where the $\BM$-th element is one and all other elements are zero. 
From the duality theorem~\cite{vanderbei1998linear}, the maximized objective function of the primal problem~(\ref{eq:Balanced}-\ref{eq:ResourceAllocation}) is equal to the minimized objective function of the dual problem~\eqref{eq:dual}: i.e., $\mu(\Income)\,{=}\,\sum_{a}I_a\hat{y}_a(\Income)$. The optimal solution $\hat{{\bf y}}$ to the dual problem thus satisfies
$\hat{y}_a\,{=}\,\partial \mu/\partial I_a$ for arbitrary $a\,({\in}\,\Exchange\,{\cup}\,\Const)$. 
To prove the monotonicity of $\hat{y}_S(I_S;\tilde{\Income})$, let us define a vector $\Income^\prime$ which differs from the vector $\Income$ only in the $S$-th element: i.e., $I^\prime_S\,{=}\,I_S\,{+}\,\Delta{I_S}$ and $I^\prime_m\,{=}\,I_m$ for $m\,({\neq}\,S)$. 
{From the definition, the solutions of the minimization problem~\eqref{eq:dual} with the parameters $\Income$ and $\Income^\prime$, $\hat{{\bf y}}(\Income)$ and $\hat{{\bf y}}(\Income^\prime)$, satisfy} 
$ \Income \cdot \hat{{\bf y}}(\Income^\prime) \geq \Income \cdot \hat{{\bf y}}(\Income) $ 
and 
$ \Income^\prime \cdot \hat{{\bf y}}(\Income) \geq \Income^\prime \cdot \hat{{\bf y}}(\Income^\prime) $. It follows
\begin{align*} 
    & \Income \cdot \hat{{\bf y}}(\Income^\prime)
    + \Income^\prime \cdot \hat{{\bf y}}(\Income) \geq 
    \Income \cdot \hat{{\bf y}}(\Income) 
    + \Income^\prime \cdot \hat{{\bf y}}(\Income^\prime) \\
    & \quad \Leftrightarrow
    \left( \Income^\prime - \Income \right)\cdot
    \left[ \hat{{\bf y}}(\Income^\prime) - \hat{{\bf y}}(\Income) \right] \leq 0\\
    & \quad \Leftrightarrow
    \Delta{I_S}
    \left[ \hat{y}_S(\Income^\prime) - \hat{y}_S(\Income) \right] \leq 0\\
    & \quad \Leftrightarrow 
     \frac{\hat{y}_S(I_S + \Delta{I_S};\tilde{\Income}) - \hat{y}_S(I_S; \tilde{\Income})}{\Delta{I_S}} \leq 0.
\end{align*}
The duality also allows us to prove the concavity of $\mu(\Income)$ as a multivariable function of $\Income$ (Appendix~\ref{sec:concavity}).

It is thus concluded that the monotonicity and concavity of the microbial growth kinetics curve, fundamental characteristics (i-ii), are universal for LP problems including CBM. 
{Moreover, our proof of monotonicity and convexity is readily generalizable to even nonlinear convex optimization problem formulations~\cite{taylor2024convex} (Appendix~\ref{sec:concavity}.1); this generalization allows for the consideration of additional factors of cellular metabolism and growth, such as the inclusion of metabolite concentrations as variables and their relationship to reaction fluxes as constraints, which are not included in standard CBM frameworks.} 

{
    Note that, the growth kinetics curve discussed in the context of Monod's law is $\mu([S])$, which argument is not maximal nutrient influx $I_S$ but nutrient concentration $[S]$. However, as long as the intake function $I_S([S])$ satisfies the monotonicity and convexity as in the Michaelis--Menten equation $I_S\,{\propto}\, [S]/(K_{\rm M}\,{+}\,[S])$ with the Michaelis constant $K_{\rm M}$ or a linear transport $I_S\,{\propto}\,[S]$, the monotonicity and convexity of the growth kinetics curve, the composite function $\mu([S])\,{=}\,\mu(I_S; \tilde{\Income})\,{\circ}\,I_S([S])$, are satisfied.}

To confirm the above analytical results, we also performed numerical simulations of {various CBM methods with genome-scale metabolic networks; they include either constraint on the allocation of proteins~\cite{mori2016constrained}, cytoplasmic volume~\cite{beg2007intracellular,vazquez2008impact}, or membrane surface area~\cite{carlson2024cell} (see Appendix~\ref{sec:CBMs} for details)}. 
The resulting piecewise linear growth kinetics curve $\mu(I_{\rm glc})$ is indeed monotonically increasing and concave, as its slope $\hat{y}_{\rm glc}\,{=}\,\partial \mu/\partial I_{\rm glc}$ monotonically decreases to zero (Fig.~\ref{fig:VariousFBAs}). 

Each locally linear part of the microbial growth kinetics curve represents a distinct phenotype phase, i.e., a qualitatively different metabolic phenotype with a different combination of growth-limiting constraints~\cite{edwards2002characterizing}. 
{In the examples of CBMs in Fig.~\ref{fig:VariousFBAs}, with small $I_{\rm glc}$}, only carbon source intake is the growth-limiting factor~\footnote{
    The region with almost zero $I_{\rm glc}$ {(for the example of CAFBA, shown in gray in Fig.~\ref{fig:general_ex})} corresponds to the cases with no solution due to the non-growth-associated maintenance energy requirements~\cite{thiele2010protocol}. It is consistent with the empirically observed minimum substrate concentration $[S]_{\min}$ to achieve cell growth.}. 
Due to the limited amount of {protein, intracellular volume, or membrane surface area (i.e., a factor in $\Const$)}, carbon metabolism at intermediate $I_{\rm glc}$ reallocates enzymes to less efficient metabolic reactions in terms of growth yield, which gradually diminishes the growth return to additional glucose intake. 
Finally, the {availability of resource(s) other than glucose} determines the maximum growth rate. 

\begin{figure}[t]
    \centering 
    \includegraphics[width=0.99\linewidth, clip]{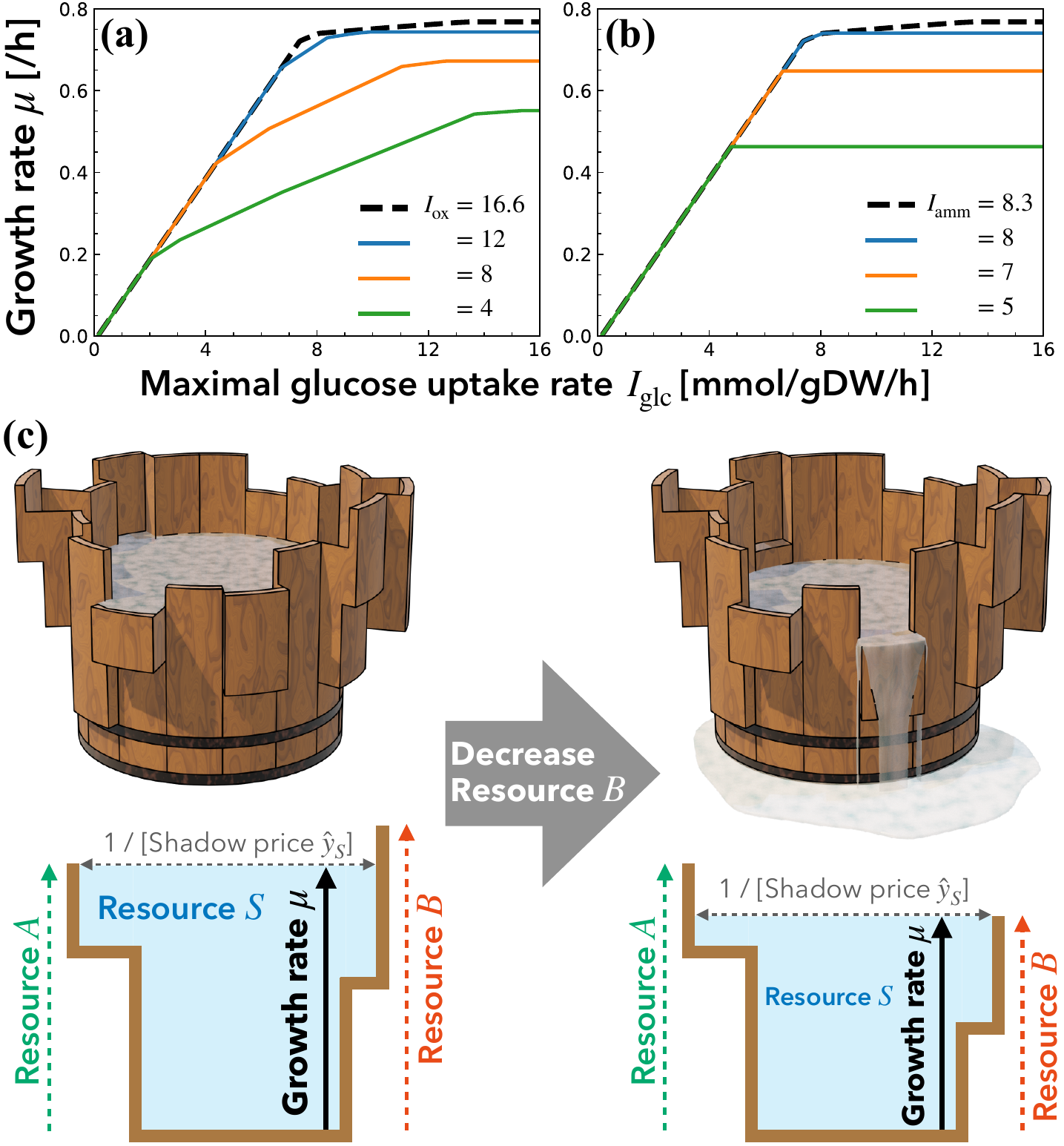}
    \caption{
        Growth rate $\mu$ as a function of carbon source availability $I_\mathrm{glc}$ with different {maximal influxes of (a) oxygen $I_{\rm ox}$ and (b) nitrogen source $I_{\rm amm}$. 
        The dashed lines correspond to the case with $I_{\rm ox}\,{=}\,16.6$ and $I_{\rm amm}\,{=}\,8.3$. 
        Numerical calculations of CAFBA~\cite{mori2016constrained} were performed using the genome-scale \textit{E. coli} iJO1366 model~\cite{orth2011comprehensive} and the COBRApy package~\cite{ebrahim2013cobrapy} (see Appendix~\ref{sec:CBMs}.1 for details). 
        } 
        {
        (c) Schematic with terraced Liebig's barrels (see also Fig.~\ref{fig:Schematics}c). 
        (left) The shortest stave corresponds to resource $A$, which has the lowest availability, and thus it determines the maximum growth rate. 
        (right) Due to a decrease in the availability of resource $B$, the maximum growth rate is limited by $B$. It leads to the reallocation of $B$ to metabolic processes and the resulting decrease in the shadow price $\hat{y}_S\,{=}\,\partial \mu/\partial I_S$ at a lower $I_S$. 
        }
    }\label{fig:general_ex2}
\end{figure}

\textit{Growth rate dependence on multiple nutrients.---}
The global constraint {principle} not only reproduces the known fundamental characteristics (i-ii) of the microbial growth kinetics curve, but also {leads to} a phenomenological theory for the dependence of growth rate on multiple nutrients. 

Since the growth kinetics curve $\mu(I_S; \tilde{\Income})$ is shaped by the global regulation of intracellular metabolism, it depends not only on the availability of the focal nutrient $S$ but also on that of the other resources, here denoted as $\tilde{\Income}$. If the {availability of a non-focal resource, i.e., a resource other than $S$,} increases (decreases), the constraint on the allocation of {the non-focal resource} is relaxed (tightened); as a result, the growth kinetics curve $\mu(I_S; \tilde{\Income})$ should shift up (down) in the phase(s) where cell growth is limited by {the non-focal resource}
(see also {Fig.~\ref{fig:general_ex2}c}). 
Conversely, by measuring how the shape of the microbial growth kinetics curve responds to environmental manipulations, such as the addition of a nutrient to the medium, one can empirically infer the growth-limiting factors under the original environmental conditions.

Such responses of the microbial growth kinetics curve are indeed observed numerically (Fig.~\ref{fig:general_ex2}). 
{The decrease in the availability of a non-focal resource, oxygen $I_{\rm ox}$ (Fig.~\ref{fig:general_ex2}a) or a nitrogen source (ammonium) $I_{\rm amm}$ (Fig.~\ref{fig:general_ex2}b), shifts the growth kinetics curve $\mu(I_{\rm glc})$ downward in a} phase with sufficiently large $I_{\rm glc}${, while maintaining the same slope $\hat{y}_S\,{=}\,\partial\mu/\partial I_S$ until the initial downward shift. Then,} one can empirically conclude that cell growth in this phase is limited by {the non-focal resource} under the original environment. 
{
    {These behaviors, which are not captured by Monod's law, can be tested experimentally.} 
    In fact, experiments using nitrogen-limited chemostats~\cite{larsson1993growth} yielded qualitatively consistent results, while further experiments are needed for thorough validation.
}

In this Letter, we have shown that the fundamental characteristics of the microbial growth kinetics curve can be explained as general properties of optimal resource allocation in cellular metabolism: as the availability of a nutrient increases, its metabolism into biomass becomes constrained by the intracellular allocation of other resources, which gradually diminishes the growth return to the additional intake of the nutrient (see also Fig.~\ref{fig:Schematics}bc). 
Note that microbial growth kinetics curves are often naively assumed to be smooth functions in the form of the Monod equation~\eqref{eq:Monod}; however, apparently multiphasic growth kinetics curves, which are non-smooth at the boundaries between different phenotype phases, have been observed experimentally~\cite{sunda2009ammonium,Heinemann-Gibbs2019,canelas2011vivo}. 
The global constraint {principle} can describe the microbial growth kinetics curve with arbitrary precision, at least if a sufficient number of constraints are considered. 

While previous studies have often argued specific (non-nutrient) constraints on the relationship between cellular growth and metabolism~\cite{yamagishi2021microeconomics,mori2016constrained,memRealEstate2,Heinemann-Gibbs2019,vazquez2008impact,beg2007intracellular,de2020common}, our approach explores a universality that is not necessarily attributable to specific molecular-biological mechanisms. 
    This approach is in line with the physics of living systems, and has been established in previous research on the ``microeconomics of metabolism''~\cite{yamagishi2021microeconomics,yamagishi2023linear}.

The global constraint {principle} for microbial growth kinetics also provides a theoretical basis for the dependence of growth rate on the availability of multiple nutrient sources, which cannot be captured by the classical arguments about Monod's bacterial growth model~{\footnote{
    {
    In Fig.~\ref{fig:general_ex2} the metabolism of non-substitutable nutrients was mainly considered. The global constraint principle is also applicable to the metabolism of substitutable nutrients, such as the mixed carbon source situations~\cite{aidelberg2014hierarchy}. 
    As discussed in ref.~\cite{aidelberg2014hierarchy}, one might naively expect that nutrients should be evolutionarily prioritized based on their contribution to growth rates.  
    Even in such situations, the monotonicity and concavity of microbial growth kinetics curves should hold.} 
    }}. 
In particular, we elucidated how the microbial growth kinetics curve $\mu(I_S)$ responds to manipulating the environmental availability of a nutrient other than the focal nutrient $S$. 
Remarkably, this model can be viewed as a generalization of Liebig's law of the minimum, which states that the growth rate of an organism is determined by the availability of the scarcest resource. 
Liebig's law is commonly illustrated by the metaphor of a barrel with uneven staves~\cite{whitson1909notes}, as the feasible water surface height of the barrel is limited by the shortest stave. This barrel metaphor describes the effect of resource availability on the maximum growth rate, but does not account for growth kinetics. 
In contrast, the global constraint {principle}, which incorporates the influence of resource allocation on growth kinetics, is also illustrated schematically using a modified Liebig's barrel with terraced staves (Fig.~\ref{fig:Schematics}c): at the water level where an additional global constraint appears, each stave of the terraced Liebig's barrel spreads out in a stepwise manner, making further increases in growth rate more difficult. 

{
    Finally, we acknowledge certain limitations. This study examines how optimized metabolic regulation universally imposes the monotonic and concave relationship between nutrient availability and growth rates. 
    Therefore, if the optimization assumption fails, for example due to the toxicity of acids and alcohols at high concentrations~\cite{wilbanks2017comprehensive}, the monotonicity may break down~\cite{bren2016glucose,wilbanks2017comprehensive}. 
    On the contrary, based on the global constraint principle, experimental deviations from monotonicity or convexity indicate additional growth-limiting factors beyond optimized metabolic regulation on cellular growth.}

The present study revisited the classical phenomenological laws in biology, Monod's law of bacterial growth and Liebig's law of the minimum, from the perspective of resource allocation in cellular metabolism. 
We {have thereby refined them into a comprehensive theory of the law of diminishing returns in biology.} 
As Liebig's law was originally formulated to describe the growth of plants, we expect our theory to provide a theoretical basis for the growth of higher organisms as well as microbes in the future. \\

We would like to thank Takuma \={O}nishi for his assistance in the preparation of Figs.~\ref{fig:Schematics}c~{and~\ref{fig:general_ex2}c and Saburo Tsuru for helpful discussions}. 
J.~F.~Y. is supported by the RIKEN Research Fund for Special Postdoctoral Researcher (project code: 202401061031) and the Masason Foundation. T.~S.~H. is supported by JSPS KAKENHI Grant Number JP21K15048.

\bibliography{bibliography}

%merlin.mbs apsrev4-1.bst 2010-07-25 4.21a (PWD, AO, DPC) hacked
%Control: key (0)
%Control: author (0) dotless jnrlst
%Control: editor formatted (1) identically to author
%Control: production of article title (0) allowed
%Control: page (1) range
%Control: year (0) verbatim
%Control: production of eprint (0) enabled
\begin{thebibliography}{60}%
\makeatletter
\providecommand \@ifxundefined [1]{%
 \@ifx{#1\undefined}
}%
\providecommand \@ifnum [1]{%
 \ifnum #1\expandafter \@firstoftwo
 \else \expandafter \@secondoftwo
 \fi
}%
\providecommand \@ifx [1]{%
 \ifx #1\expandafter \@firstoftwo
 \else \expandafter \@secondoftwo
 \fi
}%
\providecommand \natexlab [1]{#1}%
\providecommand \enquote  [1]{``#1''}%
\providecommand \bibnamefont  [1]{#1}%
\providecommand \bibfnamefont [1]{#1}%
\providecommand \citenamefont [1]{#1}%
\providecommand \href@noop [0]{\@secondoftwo}%
\providecommand \href [0]{\begingroup \@sanitize@url \@href}%
\providecommand \@href[1]{\@@startlink{#1}\@@href}%
\providecommand \@@href[1]{\endgroup#1\@@endlink}%
\providecommand \@sanitize@url [0]{\catcode `\\12\catcode `\$12\catcode `\&12\catcode `\#12\catcode `\^12\catcode `\_12\catcode `\%12\relax}%
\providecommand \@@startlink[1]{}%
\providecommand \@@endlink[0]{}%
\providecommand \url  [0]{\begingroup\@sanitize@url \@url }%
\providecommand \@url [1]{\endgroup\@href {#1}{\urlprefix }}%
\providecommand \urlprefix  [0]{URL }%
\providecommand \Eprint [0]{\href }%
\providecommand \doibase [0]{http://dx.doi.org/}%
\providecommand \selectlanguage [0]{\@gobble}%
\providecommand \bibinfo  [0]{\@secondoftwo}%
\providecommand \bibfield  [0]{\@secondoftwo}%
\providecommand \translation [1]{[#1]}%
\providecommand \BibitemOpen [0]{}%
\providecommand \bibitemStop [0]{}%
\providecommand \bibitemNoStop [0]{.\EOS\space}%
\providecommand \EOS [0]{\spacefactor3000\relax}%
\providecommand \BibitemShut  [1]{\csname bibitem#1\endcsname}%
\let\auto@bib@innerbib\@empty
%</preamble>
\bibitem [{\citenamefont {Kaneko}(2006)}]{kaneko2006life}%
  \BibitemOpen
  \bibfield  {author} {\bibinfo {author} {\bibfnamefont {K.}~\bibnamefont {Kaneko}},\ }\href@noop {} {\emph {\bibinfo {title} {Life: an introduction to complex systems biology}}}\ (\bibinfo  {publisher} {Springer},\ \bibinfo {year} {2006})\BibitemShut {NoStop}%
\bibitem [{\citenamefont {Monod}(1949)}]{monod1949growth}%
  \BibitemOpen
  \bibfield  {author} {\bibinfo {author} {\bibfnamefont {J.}~\bibnamefont {Monod}},\ }\bibfield  {title} {\enquote {\bibinfo {title} {The growth of bacterial cultures},}\ }\href@noop {} {\bibfield  {journal} {\bibinfo  {journal} {Annual review of microbiology}\ }\textbf {\bibinfo {volume} {3}},\ \bibinfo {pages} {371--394} (\bibinfo {year} {1949})}\BibitemShut {NoStop}%
\bibitem [{\citenamefont {Whitson}\ and\ \citenamefont {Walster}(1912)}]{whitson1912soils}%
  \BibitemOpen
  \bibfield  {author} {\bibinfo {author} {\bibfnamefont {A.~R.}\ \bibnamefont {Whitson}}\ and\ \bibinfo {author} {\bibfnamefont {H.~L.}\ \bibnamefont {Walster}},\ }\href@noop {} {\emph {\bibinfo {title} {Soils and soil fertility}}}\ (\bibinfo  {publisher} {Webb Publishing Company},\ \bibinfo {year} {1912})\BibitemShut {NoStop}%
\bibitem [{\citenamefont {Scott}\ and\ \citenamefont {Hwa}(2011)}]{scott2011bacterial}%
  \BibitemOpen
  \bibfield  {author} {\bibinfo {author} {\bibfnamefont {M.}~\bibnamefont {Scott}}\ and\ \bibinfo {author} {\bibfnamefont {T.}~\bibnamefont {Hwa}},\ }\bibfield  {title} {\enquote {\bibinfo {title} {Bacterial growth laws and their applications},}\ }\href@noop {} {\bibfield  {journal} {\bibinfo  {journal} {Current Opinion in Biotechnology}\ }\textbf {\bibinfo {volume} {22}},\ \bibinfo {pages} {559--565} (\bibinfo {year} {2011})}\BibitemShut {NoStop}%
\bibitem [{\citenamefont {Kov{\'a}rov{\'a}-Kovar}\ and\ \citenamefont {Egli}(1998)}]{kovarova1998growth}%
  \BibitemOpen
  \bibfield  {author} {\bibinfo {author} {\bibfnamefont {K.}~\bibnamefont {Kov{\'a}rov{\'a}-Kovar}}\ and\ \bibinfo {author} {\bibfnamefont {T.}~\bibnamefont {Egli}},\ }\bibfield  {title} {\enquote {\bibinfo {title} {Growth kinetics of suspended microbial cells: from single-substrate-controlled growth to mixed-substrate kinetics},}\ }\href@noop {} {\bibfield  {journal} {\bibinfo  {journal} {Microbiology and Molecular Biology Reviews}\ }\textbf {\bibinfo {volume} {62}},\ \bibinfo {pages} {646--666} (\bibinfo {year} {1998})}\BibitemShut {NoStop}%
\bibitem [{\citenamefont {Reich}\ and\ \citenamefont {Selkov}(1981)}]{reich1981energy}%
  \BibitemOpen
  \bibfield  {author} {\bibinfo {author} {\bibfnamefont {J.~G.}\ \bibnamefont {Reich}}\ and\ \bibinfo {author} {\bibfnamefont {E.}~\bibnamefont {Selkov}},\ }\href@noop {} {\emph {\bibinfo {title} {Energy metabolism of the cell: a theoretical treatise}}}\ (\bibinfo  {publisher} {Academic Press},\ \bibinfo {year} {1981})\BibitemShut {NoStop}%
\bibitem [{\citenamefont {Sher}\ \emph {et~al.}(2024)\citenamefont {Sher}, \citenamefont {Segr{\`e}},\ and\ \citenamefont {Follows}}]{sher2024quantitative}%
  \BibitemOpen
  \bibfield  {author} {\bibinfo {author} {\bibfnamefont {D.}~\bibnamefont {Sher}}, \bibinfo {author} {\bibfnamefont {D.}~\bibnamefont {Segr{\`e}}}, \ and\ \bibinfo {author} {\bibfnamefont {M.~J.}\ \bibnamefont {Follows}},\ }\bibfield  {title} {\enquote {\bibinfo {title} {Quantitative principles of microbial metabolism shared across scales},}\ }\href@noop {} {\bibfield  {journal} {\bibinfo  {journal} {Nature Microbiology}\ ,\ \bibinfo {pages} {1--14}} (\bibinfo {year} {2024})}\BibitemShut {NoStop}%
\bibitem [{\citenamefont {Liu}(2007)}]{liu2007overview}%
  \BibitemOpen
  \bibfield  {author} {\bibinfo {author} {\bibfnamefont {Y.}~\bibnamefont {Liu}},\ }\bibfield  {title} {\enquote {\bibinfo {title} {Overview of some theoretical approaches for derivation of the monod equation},}\ }\href@noop {} {\bibfield  {journal} {\bibinfo  {journal} {Applied microbiology and biotechnology}\ }\textbf {\bibinfo {volume} {73}},\ \bibinfo {pages} {1241--1250} (\bibinfo {year} {2007})}\BibitemShut {NoStop}%
\bibitem [{\citenamefont {De~Jong}\ \emph {et~al.}(2017)\citenamefont {De~Jong}, \citenamefont {Casagranda}, \citenamefont {Giordano}, \citenamefont {Cinquemani}, \citenamefont {Ropers}, \citenamefont {Geiselmann},\ and\ \citenamefont {Gouz{\'e}}}]{de2017mathematical}%
  \BibitemOpen
  \bibfield  {author} {\bibinfo {author} {\bibfnamefont {Hidde}\ \bibnamefont {De~Jong}}, \bibinfo {author} {\bibfnamefont {Stefano}\ \bibnamefont {Casagranda}}, \bibinfo {author} {\bibfnamefont {Nils}\ \bibnamefont {Giordano}}, \bibinfo {author} {\bibfnamefont {Eugenio}\ \bibnamefont {Cinquemani}}, \bibinfo {author} {\bibfnamefont {Delphine}\ \bibnamefont {Ropers}}, \bibinfo {author} {\bibfnamefont {Johannes}\ \bibnamefont {Geiselmann}}, \ and\ \bibinfo {author} {\bibfnamefont {Jean-Luc}\ \bibnamefont {Gouz{\'e}}},\ }\bibfield  {title} {\enquote {\bibinfo {title} {{Mathematical modelling of microbes: metabolism, gene expression and growth}},}\ }\href@noop {} {\bibfield  {journal} {\bibinfo  {journal} {Journal of The Royal Society Interface}\ }\textbf {\bibinfo {volume} {14}},\ \bibinfo {pages} {20170502} (\bibinfo {year} {2017})}\BibitemShut {NoStop}%
\bibitem [{\citenamefont {Merchuk}\ and\ \citenamefont {Asenjo}(1995)}]{merchuk1995monod}%
  \BibitemOpen
  \bibfield  {author} {\bibinfo {author} {\bibfnamefont {J.~C.}\ \bibnamefont {Merchuk}}\ and\ \bibinfo {author} {\bibfnamefont {J.~A.}\ \bibnamefont {Asenjo}},\ }\bibfield  {title} {\enquote {\bibinfo {title} {The monod equation and mass transfer},}\ }\href@noop {} {\bibfield  {journal} {\bibinfo  {journal} {Biotechnology and Bioengineering}\ }\textbf {\bibinfo {volume} {45}},\ \bibinfo {pages} {91--94} (\bibinfo {year} {1995})}\BibitemShut {NoStop}%
\bibitem [{\citenamefont {Jin}\ and\ \citenamefont {Bethke}(2003)}]{jin2003new}%
  \BibitemOpen
  \bibfield  {author} {\bibinfo {author} {\bibfnamefont {Q.}~\bibnamefont {Jin}}\ and\ \bibinfo {author} {\bibfnamefont {C.~M.}\ \bibnamefont {Bethke}},\ }\bibfield  {title} {\enquote {\bibinfo {title} {A new rate law describing microbial respiration},}\ }\href@noop {} {\bibfield  {journal} {\bibinfo  {journal} {Applied and Environmental Microbiology}\ }\textbf {\bibinfo {volume} {69}},\ \bibinfo {pages} {2340--2348} (\bibinfo {year} {2003})}\BibitemShut {NoStop}%
\bibitem [{\citenamefont {Heijnen}\ and\ \citenamefont {Romein}(1995)}]{heijnen1995derivation}%
  \BibitemOpen
  \bibfield  {author} {\bibinfo {author} {\bibfnamefont {J.~J.}\ \bibnamefont {Heijnen}}\ and\ \bibinfo {author} {\bibfnamefont {B.}~\bibnamefont {Romein}},\ }\bibfield  {title} {\enquote {\bibinfo {title} {Derivation of kinetic equations for growth on single substrates based on general properties of a simple metabolic network},}\ }\href@noop {} {\bibfield  {journal} {\bibinfo  {journal} {Biotechnology Progress}\ }\textbf {\bibinfo {volume} {11}},\ \bibinfo {pages} {712--716} (\bibinfo {year} {1995})}\BibitemShut {NoStop}%
\bibitem [{\citenamefont {Koch}(1997)}]{koch1997microbial}%
  \BibitemOpen
  \bibfield  {author} {\bibinfo {author} {\bibfnamefont {A.~L.}\ \bibnamefont {Koch}},\ }\bibfield  {title} {\enquote {\bibinfo {title} {Microbial physiology and ecology of slow growth},}\ }\href@noop {} {\bibfield  {journal} {\bibinfo  {journal} {Microbiology and Molecular Biology Reviews}\ }\textbf {\bibinfo {volume} {61}},\ \bibinfo {pages} {305--318} (\bibinfo {year} {1997})}\BibitemShut {NoStop}%
\bibitem [{\citenamefont {Goelzer}\ and\ \citenamefont {Fromion}(2011)}]{goelzer2011bacterial}%
  \BibitemOpen
  \bibfield  {author} {\bibinfo {author} {\bibfnamefont {A.}~\bibnamefont {Goelzer}}\ and\ \bibinfo {author} {\bibfnamefont {V.}~\bibnamefont {Fromion}},\ }\bibfield  {title} {\enquote {\bibinfo {title} {Bacterial growth rate reflects a bottleneck in resource allocation},}\ }\href@noop {} {\bibfield  {journal} {\bibinfo  {journal} {Biochimica et Biophysica Acta (BBA)-General Subjects}\ }\textbf {\bibinfo {volume} {1810}},\ \bibinfo {pages} {978--988} (\bibinfo {year} {2011})}\BibitemShut {NoStop}%
\bibitem [{\citenamefont {Bren}\ \emph {et~al.}(2016)\citenamefont {Bren}, \citenamefont {Park}, \citenamefont {Towbin}, \citenamefont {Dekel}, \citenamefont {Rabinowitz},\ and\ \citenamefont {Alon}}]{bren2016glucose}%
  \BibitemOpen
  \bibfield  {author} {\bibinfo {author} {\bibfnamefont {A.}~\bibnamefont {Bren}}, \bibinfo {author} {\bibfnamefont {J.~O.}\ \bibnamefont {Park}}, \bibinfo {author} {\bibfnamefont {B.~D.}\ \bibnamefont {Towbin}}, \bibinfo {author} {\bibfnamefont {E.}~\bibnamefont {Dekel}}, \bibinfo {author} {\bibfnamefont {J.~D.}\ \bibnamefont {Rabinowitz}}, \ and\ \bibinfo {author} {\bibfnamefont {U.}~\bibnamefont {Alon}},\ }\bibfield  {title} {\enquote {\bibinfo {title} {Glucose becomes one of the worst carbon sources for e. coli on poor nitrogen sources due to suboptimal levels of camp},}\ }\href@noop {} {\bibfield  {journal} {\bibinfo  {journal} {Scientific Reports}\ }\textbf {\bibinfo {volume} {6}},\ \bibinfo {pages} {24834} (\bibinfo {year} {2016})}\BibitemShut {NoStop}%
\bibitem [{\citenamefont {Tang}\ and\ \citenamefont {Riley}(2021)}]{tang2021finding}%
  \BibitemOpen
  \bibfield  {author} {\bibinfo {author} {\bibfnamefont {J.}~\bibnamefont {Tang}}\ and\ \bibinfo {author} {\bibfnamefont {W.~J.}\ \bibnamefont {Riley}},\ }\bibfield  {title} {\enquote {\bibinfo {title} {Finding liebig's law of the minimum},}\ }\href@noop {} {\bibfield  {journal} {\bibinfo  {journal} {Ecological Applications}\ }\textbf {\bibinfo {volume} {31}},\ \bibinfo {pages} {e02458} (\bibinfo {year} {2021})}\BibitemShut {NoStop}%
\bibitem [{\citenamefont {Palsson}(2015)}]{palsson2015systems}%
  \BibitemOpen
  \bibfield  {author} {\bibinfo {author} {\bibfnamefont {B.~{\O}.}\ \bibnamefont {Palsson}},\ }\href@noop {} {\emph {\bibinfo {title} {Systems Biology: Constraint-based Reconstruction and Analysis}}}\ (\bibinfo  {publisher} {Cambridge University Press},\ \bibinfo {year} {2015})\BibitemShut {NoStop}%
\bibitem [{\citenamefont {Sommer}(1991)}]{sommer1991comparison}%
  \BibitemOpen
  \bibfield  {author} {\bibinfo {author} {\bibfnamefont {U.}~\bibnamefont {Sommer}},\ }\bibfield  {title} {\enquote {\bibinfo {title} {A comparison of the droop and the monod models of nutrient limited growth applied to natural populations of phytoplankton},}\ }\href@noop {} {\bibfield  {journal} {\bibinfo  {journal} {Functional Ecology}\ ,\ \bibinfo {pages} {535--544}} (\bibinfo {year} {1991})}\BibitemShut {NoStop}%
\bibitem [{\citenamefont {Wang}\ \emph {et~al.}(2022)\citenamefont {Wang}, \citenamefont {Garcia}, \citenamefont {Ahmed},\ and\ \citenamefont {Heggerud}}]{wang2022mathematical}%
  \BibitemOpen
  \bibfield  {author} {\bibinfo {author} {\bibfnamefont {H.}~\bibnamefont {Wang}}, \bibinfo {author} {\bibfnamefont {P.~V.}\ \bibnamefont {Garcia}}, \bibinfo {author} {\bibfnamefont {S.}~\bibnamefont {Ahmed}}, \ and\ \bibinfo {author} {\bibfnamefont {C.~M.}\ \bibnamefont {Heggerud}},\ }\bibfield  {title} {\enquote {\bibinfo {title} {Mathematical comparison and empirical review of the monod and droop forms for resource-based population dynamics},}\ }\href@noop {} {\bibfield  {journal} {\bibinfo  {journal} {Ecological Modelling}\ }\textbf {\bibinfo {volume} {466}},\ \bibinfo {pages} {109887} (\bibinfo {year} {2022})}\BibitemShut {NoStop}%
\bibitem [{Note1()}]{Note1}%
  \BibitemOpen
  \bibinfo {note} {Empirically, there can be a finite minimum substrate concentration $[S]_{\min }$ to achieve cell growth due to the maintenance energy requirements~\cite {kovarova1998growth, goelzer2011bacterial}. However, in classical arguments such as Monod's, they are effectively ignored for simplicity by considering the substrate concentration $[S]$ after subtracting $[S]_{\min }$.}\BibitemShut {Stop}%
\bibitem [{\citenamefont {von Liebig}(1840)}]{von1840organic}%
  \BibitemOpen
  \bibfield  {author} {\bibinfo {author} {\bibfnamefont {J.}~\bibnamefont {von Liebig}},\ }\href@noop {} {\emph {\bibinfo {title} {Organic chemistry in its applications to agriculture and physiology}}}\ (\bibinfo  {publisher} {Taylor and Walton},\ \bibinfo {year} {1840})\BibitemShut {NoStop}%
\bibitem [{\citenamefont {Lewis}\ \emph {et~al.}(2012)\citenamefont {Lewis}, \citenamefont {Nagarajan},\ and\ \citenamefont {Palsson}}]{lewis2012constraining}%
  \BibitemOpen
  \bibfield  {author} {\bibinfo {author} {\bibfnamefont {N.~E.}\ \bibnamefont {Lewis}}, \bibinfo {author} {\bibfnamefont {H.}~\bibnamefont {Nagarajan}}, \ and\ \bibinfo {author} {\bibfnamefont {B.~{\O}.}\ \bibnamefont {Palsson}},\ }\bibfield  {title} {\enquote {\bibinfo {title} {Constraining the metabolic genotype--phenotype relationship using a phylogeny of in silico methods},}\ }\href@noop {} {\bibfield  {journal} {\bibinfo  {journal} {Nature Reviews Microbiology}\ }\textbf {\bibinfo {volume} {10}},\ \bibinfo {pages} {291--305} (\bibinfo {year} {2012})}\BibitemShut {NoStop}%
\bibitem [{\citenamefont {Ibarra}\ \emph {et~al.}(2002)\citenamefont {Ibarra}, \citenamefont {Edwards},\ and\ \citenamefont {Palsson}}]{ibarra2002escherichia}%
  \BibitemOpen
  \bibfield  {author} {\bibinfo {author} {\bibfnamefont {R.~U.}\ \bibnamefont {Ibarra}}, \bibinfo {author} {\bibfnamefont {J.~S.}\ \bibnamefont {Edwards}}, \ and\ \bibinfo {author} {\bibfnamefont {B.~{\O}.}\ \bibnamefont {Palsson}},\ }\bibfield  {title} {\enquote {\bibinfo {title} {Escherichia coli k-12 undergoes adaptive evolution to achieve in silico predicted optimal growth},}\ }\href@noop {} {\bibfield  {journal} {\bibinfo  {journal} {Nature}\ }\textbf {\bibinfo {volume} {420}},\ \bibinfo {pages} {186--189} (\bibinfo {year} {2002})}\BibitemShut {NoStop}%
\bibitem [{\citenamefont {Klipp}\ \emph {et~al.}(2016)\citenamefont {Klipp}, \citenamefont {Liebermeister}, \citenamefont {Wierling},\ and\ \citenamefont {Kowald}}]{klipp2016systems}%
  \BibitemOpen
  \bibfield  {author} {\bibinfo {author} {\bibfnamefont {E.}~\bibnamefont {Klipp}}, \bibinfo {author} {\bibfnamefont {W.}~\bibnamefont {Liebermeister}}, \bibinfo {author} {\bibfnamefont {C.}~\bibnamefont {Wierling}}, \ and\ \bibinfo {author} {\bibfnamefont {A.}~\bibnamefont {Kowald}},\ }\href@noop {} {\emph {\bibinfo {title} {Systems Biology: a textbook}}}\ (\bibinfo  {publisher} {John Wiley \& Sons},\ \bibinfo {address} {New Jersey},\ \bibinfo {year} {2016})\BibitemShut {NoStop}%
\bibitem [{\citenamefont {Warren}\ and\ \citenamefont {Jones}(2007)}]{warren2007duality}%
  \BibitemOpen
  \bibfield  {author} {\bibinfo {author} {\bibfnamefont {P.~B.}\ \bibnamefont {Warren}}\ and\ \bibinfo {author} {\bibfnamefont {J.~L.}\ \bibnamefont {Jones}},\ }\bibfield  {title} {\enquote {\bibinfo {title} {Duality, thermodynamics, and the linear programming problem in constraint-based models of metabolism},}\ }\href@noop {} {\bibfield  {journal} {\bibinfo  {journal} {Physical Review Letters}\ }\textbf {\bibinfo {volume} {99}},\ \bibinfo {pages} {108101} (\bibinfo {year} {2007})}\BibitemShut {NoStop}%
\bibitem [{\citenamefont {Yamagishi}\ and\ \citenamefont {Hatakeyama}(2023)}]{yamagishi2023linear}%
  \BibitemOpen
  \bibfield  {author} {\bibinfo {author} {\bibfnamefont {J.~F.}\ \bibnamefont {Yamagishi}}\ and\ \bibinfo {author} {\bibfnamefont {T.~S.}\ \bibnamefont {Hatakeyama}},\ }\bibfield  {title} {\enquote {\bibinfo {title} {Linear response theory of evolved metabolic systems},}\ }\href@noop {} {\bibfield  {journal} {\bibinfo  {journal} {Physical Review Letters}\ }\textbf {\bibinfo {volume} {131}},\ \bibinfo {pages} {028401} (\bibinfo {year} {2023})}\BibitemShut {NoStop}%
\bibitem [{\citenamefont {Basan}\ \emph {et~al.}(2015)\citenamefont {Basan}, \citenamefont {Hui}, \citenamefont {Okano}, \citenamefont {Zhang}, \citenamefont {Shen}, \citenamefont {Williamson},\ and\ \citenamefont {Hwa}}]{Hwa_OM}%
  \BibitemOpen
  \bibfield  {author} {\bibinfo {author} {\bibfnamefont {M.}~\bibnamefont {Basan}}, \bibinfo {author} {\bibfnamefont {S.}~\bibnamefont {Hui}}, \bibinfo {author} {\bibfnamefont {H.}~\bibnamefont {Okano}}, \bibinfo {author} {\bibfnamefont {Z.}~\bibnamefont {Zhang}}, \bibinfo {author} {\bibfnamefont {Y.}~\bibnamefont {Shen}}, \bibinfo {author} {\bibfnamefont {J.~R.}\ \bibnamefont {Williamson}}, \ and\ \bibinfo {author} {\bibfnamefont {T.}~\bibnamefont {Hwa}},\ }\bibfield  {title} {\enquote {\bibinfo {title} {Overflow metabolism in escherichia coli results from efficient proteome allocation},}\ }\href@noop {} {\bibfield  {journal} {\bibinfo  {journal} {Nature}\ }\textbf {\bibinfo {volume} {528}},\ \bibinfo {pages} {99--104} (\bibinfo {year} {2015})}\BibitemShut {NoStop}%
\bibitem [{\citenamefont {Mori}\ \emph {et~al.}(2016)\citenamefont {Mori}, \citenamefont {Hwa}, \citenamefont {Martin}, \citenamefont {De~Martino},\ and\ \citenamefont {Marinari}}]{mori2016constrained}%
  \BibitemOpen
  \bibfield  {author} {\bibinfo {author} {\bibfnamefont {M.}~\bibnamefont {Mori}}, \bibinfo {author} {\bibfnamefont {T.}~\bibnamefont {Hwa}}, \bibinfo {author} {\bibfnamefont {O.~C.}\ \bibnamefont {Martin}}, \bibinfo {author} {\bibfnamefont {A.}~\bibnamefont {De~Martino}}, \ and\ \bibinfo {author} {\bibfnamefont {E.}~\bibnamefont {Marinari}},\ }\bibfield  {title} {\enquote {\bibinfo {title} {Constrained allocation flux balance analysis},}\ }\href@noop {} {\bibfield  {journal} {\bibinfo  {journal} {PLoS Computational Biology}\ }\textbf {\bibinfo {volume} {12}},\ \bibinfo {pages} {e1004913} (\bibinfo {year} {2016})}\BibitemShut {NoStop}%
\bibitem [{\citenamefont {Szenk}\ \emph {et~al.}(2017)\citenamefont {Szenk}, \citenamefont {Dill},\ and\ \citenamefont {de~Graff}}]{memRealEstate2}%
  \BibitemOpen
  \bibfield  {author} {\bibinfo {author} {\bibfnamefont {M.}~\bibnamefont {Szenk}}, \bibinfo {author} {\bibfnamefont {K.~A.}\ \bibnamefont {Dill}}, \ and\ \bibinfo {author} {\bibfnamefont {A.~M.~R.}\ \bibnamefont {de~Graff}},\ }\bibfield  {title} {\enquote {\bibinfo {title} {Why do fast-growing bacteria enter overflow metabolism? testing the membrane real estate hypothesis},}\ }\href@noop {} {\bibfield  {journal} {\bibinfo  {journal} {Cell Systems}\ }\textbf {\bibinfo {volume} {5}},\ \bibinfo {pages} {95--104} (\bibinfo {year} {2017})}\BibitemShut {NoStop}%
\bibitem [{\citenamefont {Niebel}\ \emph {et~al.}(2019)\citenamefont {Niebel}, \citenamefont {Leupold},\ and\ \citenamefont {Heinemann}}]{Heinemann-Gibbs2019}%
  \BibitemOpen
  \bibfield  {author} {\bibinfo {author} {\bibfnamefont {B.}~\bibnamefont {Niebel}}, \bibinfo {author} {\bibfnamefont {S.}~\bibnamefont {Leupold}}, \ and\ \bibinfo {author} {\bibfnamefont {M.}~\bibnamefont {Heinemann}},\ }\bibfield  {title} {\enquote {\bibinfo {title} {An upper limit on gibbs energy dissipation governs cellular metabolism},}\ }\href@noop {} {\bibfield  {journal} {\bibinfo  {journal} {Nature Metabolism}\ }\textbf {\bibinfo {volume} {1}},\ \bibinfo {pages} {125--132} (\bibinfo {year} {2019})}\BibitemShut {NoStop}%
\bibitem [{\citenamefont {Beg}\ \emph {et~al.}(2007)\citenamefont {Beg}, \citenamefont {Vazquez}, \citenamefont {Ernst}, \citenamefont {de~Menezes}, \citenamefont {Bar-Joseph}, \citenamefont {Barab{\'a}si},\ and\ \citenamefont {Oltvai}}]{beg2007intracellular}%
  \BibitemOpen
  \bibfield  {author} {\bibinfo {author} {\bibfnamefont {Qasim~K}\ \bibnamefont {Beg}}, \bibinfo {author} {\bibfnamefont {Alexei}\ \bibnamefont {Vazquez}}, \bibinfo {author} {\bibfnamefont {Jason}\ \bibnamefont {Ernst}}, \bibinfo {author} {\bibfnamefont {Marcio~A}\ \bibnamefont {de~Menezes}}, \bibinfo {author} {\bibfnamefont {Ziv}\ \bibnamefont {Bar-Joseph}}, \bibinfo {author} {\bibfnamefont {A-L}\ \bibnamefont {Barab{\'a}si}}, \ and\ \bibinfo {author} {\bibfnamefont {Zolt{\'a}n~N}\ \bibnamefont {Oltvai}},\ }\bibfield  {title} {\enquote {\bibinfo {title} {{Intracellular crowding defines the mode and sequence of substrate uptake by Escherichia coli and constrains its metabolic activity}},}\ }\href@noop {} {\bibfield  {journal} {\bibinfo  {journal} {Proceedings of the National Academy of Sciences}\ }\textbf {\bibinfo {volume} {104}},\ \bibinfo {pages} {12663--12668} (\bibinfo {year} {2007})}\BibitemShut {NoStop}%
\bibitem [{\citenamefont {Vazquez}\ \emph {et~al.}(2008)\citenamefont {Vazquez}, \citenamefont {Beg}, \citenamefont {DeMenezes}, \citenamefont {Ernst}, \citenamefont {Bar-Joseph}, \citenamefont {Barab{\'a}si}, \citenamefont {Boros},\ and\ \citenamefont {Oltvai}}]{vazquez2008impact}%
  \BibitemOpen
  \bibfield  {author} {\bibinfo {author} {\bibfnamefont {Alexei}\ \bibnamefont {Vazquez}}, \bibinfo {author} {\bibfnamefont {Qasim~K}\ \bibnamefont {Beg}}, \bibinfo {author} {\bibfnamefont {Marcio~A}\ \bibnamefont {DeMenezes}}, \bibinfo {author} {\bibfnamefont {Jason}\ \bibnamefont {Ernst}}, \bibinfo {author} {\bibfnamefont {Ziv}\ \bibnamefont {Bar-Joseph}}, \bibinfo {author} {\bibfnamefont {Albert-L{\'a}szl{\'o}}\ \bibnamefont {Barab{\'a}si}}, \bibinfo {author} {\bibfnamefont {L{\'a}szl{\'o}~G}\ \bibnamefont {Boros}}, \ and\ \bibinfo {author} {\bibfnamefont {Zolt{\'a}n~N}\ \bibnamefont {Oltvai}},\ }\bibfield  {title} {\enquote {\bibinfo {title} {{Impact of the solvent capacity constraint on E. coli metabolism}},}\ }\href@noop {} {\bibfield  {journal} {\bibinfo  {journal} {BMC systems biology}\ }\textbf {\bibinfo {volume} {2}},\ \bibinfo {pages} {1--10} (\bibinfo {year} {2008})}\BibitemShut {NoStop}%
\bibitem [{\citenamefont {Carlson}\ \emph {et~al.}(2024)\citenamefont {Carlson}, \citenamefont {Beck}, \citenamefont {Benitez}, \citenamefont {Harcombe}, \citenamefont {Mahadevan},\ and\ \citenamefont {Gedeon}}]{carlson2024cell}%
  \BibitemOpen
  \bibfield  {author} {\bibinfo {author} {\bibfnamefont {Ross~P}\ \bibnamefont {Carlson}}, \bibinfo {author} {\bibfnamefont {Ashley~E}\ \bibnamefont {Beck}}, \bibinfo {author} {\bibfnamefont {Mauricio~Garcia}\ \bibnamefont {Benitez}}, \bibinfo {author} {\bibfnamefont {William~R}\ \bibnamefont {Harcombe}}, \bibinfo {author} {\bibfnamefont {Radhakrishnan}\ \bibnamefont {Mahadevan}}, \ and\ \bibinfo {author} {\bibfnamefont {Tom{\'a}{\v{s}}}\ \bibnamefont {Gedeon}},\ }\bibfield  {title} {\enquote {\bibinfo {title} {{Cell Geometry and Membrane Protein Crowding Constrain Growth Rate, Overflow Metabolism, Respiration, and Maintenance Energy}},}\ }\href@noop {} {\bibfield  {journal} {\bibinfo  {journal} {bioRxiv}\ } (\bibinfo {year} {2024})}\BibitemShut {NoStop}%
\bibitem [{\citenamefont {Goelzer}\ \emph {et~al.}(2011)\citenamefont {Goelzer}, \citenamefont {Fromion},\ and\ \citenamefont {Scorletti}}]{goelzer2011cell}%
  \BibitemOpen
  \bibfield  {author} {\bibinfo {author} {\bibfnamefont {Anne}\ \bibnamefont {Goelzer}}, \bibinfo {author} {\bibfnamefont {Vincent}\ \bibnamefont {Fromion}}, \ and\ \bibinfo {author} {\bibfnamefont {G{\'e}rard}\ \bibnamefont {Scorletti}},\ }\bibfield  {title} {\enquote {\bibinfo {title} {Cell design in bacteria as a convex optimization problem},}\ }\href@noop {} {\bibfield  {journal} {\bibinfo  {journal} {Automatica}\ }\textbf {\bibinfo {volume} {47}},\ \bibinfo {pages} {1210--1218} (\bibinfo {year} {2011})}\BibitemShut {NoStop}%
\bibitem [{\citenamefont {Shinfuku}\ \emph {et~al.}(2009)\citenamefont {Shinfuku}, \citenamefont {Sorpitiporn}, \citenamefont {Sono}, \citenamefont {Furusawa}, \citenamefont {Hirasawa},\ and\ \citenamefont {Shimizu}}]{shinfuku2009development}%
  \BibitemOpen
  \bibfield  {author} {\bibinfo {author} {\bibfnamefont {Y.}~\bibnamefont {Shinfuku}}, \bibinfo {author} {\bibfnamefont {N.}~\bibnamefont {Sorpitiporn}}, \bibinfo {author} {\bibfnamefont {M.}~\bibnamefont {Sono}}, \bibinfo {author} {\bibfnamefont {C.}~\bibnamefont {Furusawa}}, \bibinfo {author} {\bibfnamefont {T.}~\bibnamefont {Hirasawa}}, \ and\ \bibinfo {author} {\bibfnamefont {H.}~\bibnamefont {Shimizu}},\ }\bibfield  {title} {\enquote {\bibinfo {title} {Development and experimental verification of a genome-scale metabolic model for corynebacterium glutamicum},}\ }\href@noop {} {\bibfield  {journal} {\bibinfo  {journal} {Microbial Cell Factories}\ }\textbf {\bibinfo {volume} {8}},\ \bibinfo {pages} {1--15} (\bibinfo {year} {2009})}\BibitemShut {NoStop}%
\bibitem [{\citenamefont {Zeng}\ and\ \citenamefont {Yang}(2020)}]{zeng2020bridging}%
  \BibitemOpen
  \bibfield  {author} {\bibinfo {author} {\bibfnamefont {H.}~\bibnamefont {Zeng}}\ and\ \bibinfo {author} {\bibfnamefont {A.}~\bibnamefont {Yang}},\ }\bibfield  {title} {\enquote {\bibinfo {title} {Bridging substrate intake kinetics and bacterial growth phenotypes with flux balance analysis incorporating proteome allocation},}\ }\href@noop {} {\bibfield  {journal} {\bibinfo  {journal} {Scientific Reports}\ }\textbf {\bibinfo {volume} {10}},\ \bibinfo {pages} {4283} (\bibinfo {year} {2020})}\BibitemShut {NoStop}%
\bibitem [{\citenamefont {Yamagishi}\ and\ \citenamefont {Hatakeyama}(2021)}]{yamagishi2021microeconomics}%
  \BibitemOpen
  \bibfield  {author} {\bibinfo {author} {\bibfnamefont {J.~F.}\ \bibnamefont {Yamagishi}}\ and\ \bibinfo {author} {\bibfnamefont {T.~S.}\ \bibnamefont {Hatakeyama}},\ }\bibfield  {title} {\enquote {\bibinfo {title} {Microeconomics of metabolism: the warburg effect as giffen behaviour},}\ }\href@noop {} {\bibfield  {journal} {\bibinfo  {journal} {Bulletin of Mathematical Biology}\ }\textbf {\bibinfo {volume} {83}},\ \bibinfo {pages} {120} (\bibinfo {year} {2021})}\BibitemShut {NoStop}%
\bibitem [{\citenamefont {Reznik}\ \emph {et~al.}(2013)\citenamefont {Reznik}, \citenamefont {Mehta},\ and\ \citenamefont {Segr{\`e}}}]{reznik2013flux}%
  \BibitemOpen
  \bibfield  {author} {\bibinfo {author} {\bibfnamefont {E.}~\bibnamefont {Reznik}}, \bibinfo {author} {\bibfnamefont {P.}~\bibnamefont {Mehta}}, \ and\ \bibinfo {author} {\bibfnamefont {D.}~\bibnamefont {Segr{\`e}}},\ }\bibfield  {title} {\enquote {\bibinfo {title} {Flux imbalance analysis and the sensitivity of cellular growth to changes in metabolite pools},}\ }\href@noop {} {\bibfield  {journal} {\bibinfo  {journal} {PLoS Computational Biology}\ }\textbf {\bibinfo {volume} {9}},\ \bibinfo {pages} {e1003195} (\bibinfo {year} {2013})}\BibitemShut {NoStop}%
\bibitem [{\citenamefont {Varma}\ \emph {et~al.}(1993)\citenamefont {Varma}, \citenamefont {Boesch},\ and\ \citenamefont {Palsson}}]{varma1993stoichiometric}%
  \BibitemOpen
  \bibfield  {author} {\bibinfo {author} {\bibfnamefont {A.}~\bibnamefont {Varma}}, \bibinfo {author} {\bibfnamefont {B.~W.}\ \bibnamefont {Boesch}}, \ and\ \bibinfo {author} {\bibfnamefont {B.~{\O}.}\ \bibnamefont {Palsson}},\ }\bibfield  {title} {\enquote {\bibinfo {title} {Stoichiometric interpretation of escherichia coli glucose catabolism under various oxygenation rates},}\ }\href@noop {} {\bibfield  {journal} {\bibinfo  {journal} {Applied and environmental microbiology}\ }\textbf {\bibinfo {volume} {59}},\ \bibinfo {pages} {2465--2473} (\bibinfo {year} {1993})}\BibitemShut {NoStop}%
\bibitem [{\citenamefont {Edwards}\ \emph {et~al.}(2002)\citenamefont {Edwards}, \citenamefont {Ramakrishna},\ and\ \citenamefont {Palsson}}]{edwards2002characterizing}%
  \BibitemOpen
  \bibfield  {author} {\bibinfo {author} {\bibfnamefont {J.~S.}\ \bibnamefont {Edwards}}, \bibinfo {author} {\bibfnamefont {R.}~\bibnamefont {Ramakrishna}}, \ and\ \bibinfo {author} {\bibfnamefont {B.~{\O}.}\ \bibnamefont {Palsson}},\ }\bibfield  {title} {\enquote {\bibinfo {title} {Characterizing the metabolic phenotype: a phenotype phase plane analysis},}\ }\href@noop {} {\bibfield  {journal} {\bibinfo  {journal} {Biotechnology and Bioengineering}\ }\textbf {\bibinfo {volume} {77}},\ \bibinfo {pages} {27--36} (\bibinfo {year} {2002})}\BibitemShut {NoStop}%
\bibitem [{\citenamefont {Vanderbei}(1998)}]{vanderbei1998linear}%
  \BibitemOpen
  \bibfield  {author} {\bibinfo {author} {\bibfnamefont {R.~J.}\ \bibnamefont {Vanderbei}},\ }\bibfield  {title} {\enquote {\bibinfo {title} {Linear programming: foundations and extensions},}\ }\href@noop {} {\bibfield  {journal} {\bibinfo  {journal} {Journal of the Operational Research Society}\ }\textbf {\bibinfo {volume} {49}},\ \bibinfo {pages} {94--94} (\bibinfo {year} {1998})}\BibitemShut {NoStop}%
\bibitem [{\citenamefont {Taylor}\ \emph {et~al.}(2024)\citenamefont {Taylor}, \citenamefont {Rapaport},\ and\ \citenamefont {Dochain}}]{taylor2024convex}%
  \BibitemOpen
  \bibfield  {author} {\bibinfo {author} {\bibfnamefont {Josh~A}\ \bibnamefont {Taylor}}, \bibinfo {author} {\bibfnamefont {Alain}\ \bibnamefont {Rapaport}}, \ and\ \bibinfo {author} {\bibfnamefont {Denis}\ \bibnamefont {Dochain}},\ }\bibfield  {title} {\enquote {\bibinfo {title} {{Convex Representation of Metabolic Networks with Michaelis--Menten Kinetics}},}\ }\href@noop {} {\bibfield  {journal} {\bibinfo  {journal} {Bulletin of Mathematical Biology}\ }\textbf {\bibinfo {volume} {86}},\ \bibinfo {pages} {65} (\bibinfo {year} {2024})}\BibitemShut {NoStop}%
\bibitem [{Note2()}]{Note2}%
  \BibitemOpen
  \bibinfo {note} {The region with almost zero $I_{\protect \rm glc}$ {(for the example of CAFBA, shown in gray in Fig.~\ref {fig:general_ex})} corresponds to the cases with no solution due to the non-growth-associated maintenance energy requirements~\cite {thiele2010protocol}. It is consistent with the empirically observed minimum substrate concentration $[S]_{\min }$ to achieve cell growth.}\BibitemShut {Stop}%
\bibitem [{\citenamefont {Orth}\ \emph {et~al.}(2011)\citenamefont {Orth}, \citenamefont {Conrad}, \citenamefont {Na}, \citenamefont {Lerman}, \citenamefont {Nam}, \citenamefont {Feist},\ and\ \citenamefont {Palsson}}]{orth2011comprehensive}%
  \BibitemOpen
  \bibfield  {author} {\bibinfo {author} {\bibfnamefont {J.~D.}\ \bibnamefont {Orth}}, \bibinfo {author} {\bibfnamefont {T.~M.}\ \bibnamefont {Conrad}}, \bibinfo {author} {\bibfnamefont {J.}~\bibnamefont {Na}}, \bibinfo {author} {\bibfnamefont {J.~A.}\ \bibnamefont {Lerman}}, \bibinfo {author} {\bibfnamefont {H.}~\bibnamefont {Nam}}, \bibinfo {author} {\bibfnamefont {A.~M.}\ \bibnamefont {Feist}}, \ and\ \bibinfo {author} {\bibfnamefont {B.~{\O}.}\ \bibnamefont {Palsson}},\ }\bibfield  {title} {\enquote {\bibinfo {title} {A comprehensive genome-scale reconstruction of escherichia coli metabolism―2011},}\ }\href@noop {} {\bibfield  {journal} {\bibinfo  {journal} {Molecular Systems Biology}\ }\textbf {\bibinfo {volume} {7}},\ \bibinfo {pages} {535} (\bibinfo {year} {2011})}\BibitemShut {NoStop}%
\bibitem [{\citenamefont {Ebrahim}\ \emph {et~al.}(2013)\citenamefont {Ebrahim}, \citenamefont {Lerman}, \citenamefont {Palsson},\ and\ \citenamefont {Hyduke}}]{ebrahim2013cobrapy}%
  \BibitemOpen
  \bibfield  {author} {\bibinfo {author} {\bibfnamefont {A.}~\bibnamefont {Ebrahim}}, \bibinfo {author} {\bibfnamefont {J.~A.}\ \bibnamefont {Lerman}}, \bibinfo {author} {\bibfnamefont {B.~{\O}.}\ \bibnamefont {Palsson}}, \ and\ \bibinfo {author} {\bibfnamefont {D.~R.}\ \bibnamefont {Hyduke}},\ }\bibfield  {title} {\enquote {\bibinfo {title} {Cobrapy: Constraints-based reconstruction and analysis for python},}\ }\href@noop {} {\bibfield  {journal} {\bibinfo  {journal} {BMC Syst Biol}\ }\textbf {\bibinfo {volume} {7}} (\bibinfo {year} {2013})}\BibitemShut {NoStop}%
\bibitem [{\citenamefont {Larsson}\ \emph {et~al.}(1993)\citenamefont {Larsson}, \citenamefont {von Stockar}, \citenamefont {Marison},\ and\ \citenamefont {Gustafsson}}]{larsson1993growth}%
  \BibitemOpen
  \bibfield  {author} {\bibinfo {author} {\bibfnamefont {Christer}\ \bibnamefont {Larsson}}, \bibinfo {author} {\bibfnamefont {Urs}\ \bibnamefont {von Stockar}}, \bibinfo {author} {\bibfnamefont {Ian}\ \bibnamefont {Marison}}, \ and\ \bibinfo {author} {\bibfnamefont {Lena}\ \bibnamefont {Gustafsson}},\ }\bibfield  {title} {\enquote {\bibinfo {title} {Growth and metabolism of saccharomyces cerevisiae in chemostat cultures under carbon-, nitrogen-, or carbon-and nitrogen-limiting conditions},}\ }\href@noop {} {\bibfield  {journal} {\bibinfo  {journal} {Journal of bacteriology}\ }\textbf {\bibinfo {volume} {175}},\ \bibinfo {pages} {4809--4816} (\bibinfo {year} {1993})}\BibitemShut {NoStop}%
\bibitem [{\citenamefont {Sunda}\ \emph {et~al.}(2009)\citenamefont {Sunda}, \citenamefont {Shertzer},\ and\ \citenamefont {Hardison}}]{sunda2009ammonium}%
  \BibitemOpen
  \bibfield  {author} {\bibinfo {author} {\bibfnamefont {W.~G.}\ \bibnamefont {Sunda}}, \bibinfo {author} {\bibfnamefont {K.~W.}\ \bibnamefont {Shertzer}}, \ and\ \bibinfo {author} {\bibfnamefont {D.~R.}\ \bibnamefont {Hardison}},\ }\bibfield  {title} {\enquote {\bibinfo {title} {Ammonium uptake and growth models in marine diatoms: Monod and droop revisited},}\ }\href@noop {} {\bibfield  {journal} {\bibinfo  {journal} {Marine Ecology Progress Series}\ }\textbf {\bibinfo {volume} {386}},\ \bibinfo {pages} {29--41} (\bibinfo {year} {2009})}\BibitemShut {NoStop}%
\bibitem [{\citenamefont {Canelas}\ \emph {et~al.}(2011)\citenamefont {Canelas}, \citenamefont {Ras}, \citenamefont {ten Pierick}, \citenamefont {van Gulik},\ and\ \citenamefont {Heijnen}}]{canelas2011vivo}%
  \BibitemOpen
  \bibfield  {author} {\bibinfo {author} {\bibfnamefont {Andr{\'e}~B}\ \bibnamefont {Canelas}}, \bibinfo {author} {\bibfnamefont {Cor}\ \bibnamefont {Ras}}, \bibinfo {author} {\bibfnamefont {Angela}\ \bibnamefont {ten Pierick}}, \bibinfo {author} {\bibfnamefont {Walter~M}\ \bibnamefont {van Gulik}}, \ and\ \bibinfo {author} {\bibfnamefont {Joseph~J}\ \bibnamefont {Heijnen}},\ }\bibfield  {title} {\enquote {\bibinfo {title} {{An in vivo data-driven framework for classification and quantification of enzyme kinetics and determination of apparent thermodynamic data}},}\ }\href@noop {} {\bibfield  {journal} {\bibinfo  {journal} {Metabolic engineering}\ }\textbf {\bibinfo {volume} {13}},\ \bibinfo {pages} {294--306} (\bibinfo {year} {2011})}\BibitemShut {NoStop}%
\bibitem [{\citenamefont {De~Groot}\ \emph {et~al.}(2020)\citenamefont {De~Groot}, \citenamefont {Lischke}, \citenamefont {Muolo}, \citenamefont {Planqu{\'e}}, \citenamefont {Bruggeman},\ and\ \citenamefont {Teusink}}]{de2020common}%
  \BibitemOpen
  \bibfield  {author} {\bibinfo {author} {\bibfnamefont {Daan~H}\ \bibnamefont {De~Groot}}, \bibinfo {author} {\bibfnamefont {Julia}\ \bibnamefont {Lischke}}, \bibinfo {author} {\bibfnamefont {Riccardo}\ \bibnamefont {Muolo}}, \bibinfo {author} {\bibfnamefont {Robert}\ \bibnamefont {Planqu{\'e}}}, \bibinfo {author} {\bibfnamefont {Frank~J}\ \bibnamefont {Bruggeman}}, \ and\ \bibinfo {author} {\bibfnamefont {Bas}\ \bibnamefont {Teusink}},\ }\bibfield  {title} {\enquote {\bibinfo {title} {{The common message of constraint-based optimization approaches: overflow metabolism is caused by two growth-limiting constraints}},}\ }\href@noop {} {\bibfield  {journal} {\bibinfo  {journal} {Cellular and Molecular Life Sciences}\ }\textbf {\bibinfo {volume} {77}},\ \bibinfo {pages} {441--453} (\bibinfo {year} {2020})}\BibitemShut {NoStop}%
\bibitem [{Note3()}]{Note3}%
  \BibitemOpen
  \bibinfo {note} {{ In Fig.~\ref {fig:general_ex2} the metabolism of non-substitutable nutrients was mainly considered. The global constraint principle is also applicable to the metabolism of substitutable nutrients, such as the mixed carbon source situations~\cite {aidelberg2014hierarchy}. As discussed in ref.~\cite {aidelberg2014hierarchy}, one might naively expect that nutrients should be evolutionarily prioritized based on their contribution to growth rates. Even in such situations, the monotonicity and concavity of microbial growth kinetics curves should hold.}}\BibitemShut {Stop}%
\bibitem [{\citenamefont {Whitson}\ and\ \citenamefont {Walster}(1909)}]{whitson1909notes}%
  \BibitemOpen
  \bibfield  {author} {\bibinfo {author} {\bibfnamefont {A.~R.}\ \bibnamefont {Whitson}}\ and\ \bibinfo {author} {\bibfnamefont {H.~L.}\ \bibnamefont {Walster}},\ }\href@noop {} {\emph {\bibinfo {title} {Notes on Soils: An Outline for an Elementary Course in Soils}}}\ (\bibinfo  {publisher} {The authors},\ \bibinfo {year} {1909})\BibitemShut {NoStop}%
\bibitem [{\citenamefont {Wilbanks}\ and\ \citenamefont {Trinh}(2017)}]{wilbanks2017comprehensive}%
  \BibitemOpen
  \bibfield  {author} {\bibinfo {author} {\bibfnamefont {Brandon}\ \bibnamefont {Wilbanks}}\ and\ \bibinfo {author} {\bibfnamefont {Cong~T}\ \bibnamefont {Trinh}},\ }\bibfield  {title} {\enquote {\bibinfo {title} {{Comprehensive characterization of toxicity of fermentative metabolites on microbial growth}},}\ }\href@noop {} {\bibfield  {journal} {\bibinfo  {journal} {Biotechnology for biofuels}\ }\textbf {\bibinfo {volume} {10}},\ \bibinfo {pages} {1--11} (\bibinfo {year} {2017})}\BibitemShut {NoStop}%
\bibitem [{\citenamefont {Thiele}\ and\ \citenamefont {Palsson}(2010)}]{thiele2010protocol}%
  \BibitemOpen
  \bibfield  {author} {\bibinfo {author} {\bibfnamefont {I.}~\bibnamefont {Thiele}}\ and\ \bibinfo {author} {\bibfnamefont {B.~{\O}.}\ \bibnamefont {Palsson}},\ }\bibfield  {title} {\enquote {\bibinfo {title} {A protocol for generating a high-quality genome-scale metabolic reconstruction},}\ }\href@noop {} {\bibfield  {journal} {\bibinfo  {journal} {Nature Protocols}\ }\textbf {\bibinfo {volume} {5}},\ \bibinfo {pages} {93--121} (\bibinfo {year} {2010})}\BibitemShut {NoStop}%
\bibitem [{\citenamefont {Aidelberg}\ \emph {et~al.}(2014)\citenamefont {Aidelberg}, \citenamefont {Towbin}, \citenamefont {Rothschild}, \citenamefont {Dekel}, \citenamefont {Bren},\ and\ \citenamefont {Alon}}]{aidelberg2014hierarchy}%
  \BibitemOpen
  \bibfield  {author} {\bibinfo {author} {\bibfnamefont {Guy}\ \bibnamefont {Aidelberg}}, \bibinfo {author} {\bibfnamefont {Benjamin~D}\ \bibnamefont {Towbin}}, \bibinfo {author} {\bibfnamefont {Daphna}\ \bibnamefont {Rothschild}}, \bibinfo {author} {\bibfnamefont {Erez}\ \bibnamefont {Dekel}}, \bibinfo {author} {\bibfnamefont {Anat}\ \bibnamefont {Bren}}, \ and\ \bibinfo {author} {\bibfnamefont {Uri}\ \bibnamefont {Alon}},\ }\bibfield  {title} {\enquote {\bibinfo {title} {{Hierarchy of non-glucose sugars in Escherichia coli}},}\ }\href@noop {} {\bibfield  {journal} {\bibinfo  {journal} {BMC systems biology}\ }\textbf {\bibinfo {volume} {8}},\ \bibinfo {pages} {1--12} (\bibinfo {year} {2014})}\BibitemShut {NoStop}%
\bibitem [{\citenamefont {Flamholz}\ \emph {et~al.}(2013)\citenamefont {Flamholz}, \citenamefont {Noor}, \citenamefont {Bar-Even}, \citenamefont {Liebermeister},\ and\ \citenamefont {Milo}}]{EMP/ED}%
  \BibitemOpen
  \bibfield  {author} {\bibinfo {author} {\bibfnamefont {Avi}\ \bibnamefont {Flamholz}}, \bibinfo {author} {\bibfnamefont {Elad}\ \bibnamefont {Noor}}, \bibinfo {author} {\bibfnamefont {Arren}\ \bibnamefont {Bar-Even}}, \bibinfo {author} {\bibfnamefont {Wolfram}\ \bibnamefont {Liebermeister}}, \ and\ \bibinfo {author} {\bibfnamefont {Ron}\ \bibnamefont {Milo}},\ }\bibfield  {title} {\enquote {\bibinfo {title} {{Glycolytic strategy as a tradeoff between energy yield and protein cost}},}\ }\href@noop {} {\bibfield  {journal} {\bibinfo  {journal} {Proceedings of the National Academy of Sciences}\ }\textbf {\bibinfo {volume} {110}},\ \bibinfo {pages} {10039--10044} (\bibinfo {year} {2013})}\BibitemShut {NoStop}%
\bibitem [{\citenamefont {Vazquez}\ \emph {et~al.}(2010)\citenamefont {Vazquez}, \citenamefont {Liu}, \citenamefont {Zhou},\ and\ \citenamefont {Oltvai}}]{vazquez2010catabolic}%
  \BibitemOpen
  \bibfield  {author} {\bibinfo {author} {\bibfnamefont {Alexei}\ \bibnamefont {Vazquez}}, \bibinfo {author} {\bibfnamefont {Jiangxia}\ \bibnamefont {Liu}}, \bibinfo {author} {\bibfnamefont {Yi}~\bibnamefont {Zhou}}, \ and\ \bibinfo {author} {\bibfnamefont {Zolt{\'a}n~N}\ \bibnamefont {Oltvai}},\ }\bibfield  {title} {\enquote {\bibinfo {title} {Catabolic efficiency of aerobic glycolysis: the warburg effect revisited},}\ }\href@noop {} {\bibfield  {journal} {\bibinfo  {journal} {BMC Systems Biology}\ }\textbf {\bibinfo {volume} {4}},\ \bibinfo {pages} {1--9} (\bibinfo {year} {2010})}\BibitemShut {NoStop}%
\bibitem [{\citenamefont {Vazquez}(2017)}]{OMbook}%
  \BibitemOpen
  \bibfield  {author} {\bibinfo {author} {\bibfnamefont {Alexei}\ \bibnamefont {Vazquez}},\ }\href@noop {} {\emph {\bibinfo {title} {Overflow metabolism: from yeast to marathon runners}}}\ (\bibinfo  {publisher} {Academic Press},\ \bibinfo {address} {London},\ \bibinfo {year} {2017})\BibitemShut {NoStop}%
\bibitem [{\citenamefont {Taylor}\ and\ \citenamefont {Rapaport}(2021)}]{taylor2021second}%
  \BibitemOpen
  \bibfield  {author} {\bibinfo {author} {\bibfnamefont {Josh~A}\ \bibnamefont {Taylor}}\ and\ \bibinfo {author} {\bibfnamefont {Alain}\ \bibnamefont {Rapaport}},\ }\bibfield  {title} {\enquote {\bibinfo {title} {Second-order cone optimization of the gradostat},}\ }\href@noop {} {\bibfield  {journal} {\bibinfo  {journal} {Computers \& Chemical Engineering}\ }\textbf {\bibinfo {volume} {151}},\ \bibinfo {pages} {107347} (\bibinfo {year} {2021})}\BibitemShut {NoStop}%
\bibitem [{\citenamefont {Luenberger}\ \emph {et~al.}(1984)\citenamefont {Luenberger}, \citenamefont {Ye} \emph {et~al.}}]{luenberger1984linear}%
  \BibitemOpen
  \bibfield  {author} {\bibinfo {author} {\bibfnamefont {David~G}\ \bibnamefont {Luenberger}}, \bibinfo {author} {\bibfnamefont {Yinyu}\ \bibnamefont {Ye}},  \emph {et~al.},\ }\href@noop {} {\emph {\bibinfo {title} {Linear and nonlinear programming}}},\ Vol.~\bibinfo {volume} {2}\ (\bibinfo  {publisher} {Springer},\ \bibinfo {year} {1984})\BibitemShut {NoStop}%
\bibitem [{\citenamefont {Takano}\ \emph {et~al.}(2023)\citenamefont {Takano}, \citenamefont {Vila}, \citenamefont {Miyazaki}, \citenamefont {S{\'a}nchez},\ and\ \citenamefont {Baji{\'c}}}]{takano2023architecture}%
  \BibitemOpen
  \bibfield  {author} {\bibinfo {author} {\bibfnamefont {Sotaro}\ \bibnamefont {Takano}}, \bibinfo {author} {\bibfnamefont {Jean~CC}\ \bibnamefont {Vila}}, \bibinfo {author} {\bibfnamefont {Ryo}\ \bibnamefont {Miyazaki}}, \bibinfo {author} {\bibfnamefont {{\'A}lvaro}\ \bibnamefont {S{\'a}nchez}}, \ and\ \bibinfo {author} {\bibfnamefont {Djordje}\ \bibnamefont {Baji{\'c}}},\ }\bibfield  {title} {\enquote {\bibinfo {title} {The architecture of metabolic networks constrains the evolution of microbial resource hierarchies},}\ }\href@noop {} {\bibfield  {journal} {\bibinfo  {journal} {Molecular Biology and Evolution}\ }\textbf {\bibinfo {volume} {40}},\ \bibinfo {pages} {msad187} (\bibinfo {year} {2023})}\BibitemShut {NoStop}%
\end{thebibliography}%

\clearpage
\appendix
\setcounter{figure}{0}
\setcounter{table}{0}
\setcounter{page}{1}
\renewcommand{\thesubsection}{\Alph{section}.\arabic{subsection}}
\renewcommand{\thefigure}{S\arabic{figure}}
\renewcommand{\thetable}{S\arabic{table}}

\onecolumngrid
\begin{center}
    {\large\textbf{Supplemental Material} for
    Jumpei F. Yamagishi and Tetsuhiro S. Hatakeyama \\\textbf{``Global Constraint {Principle} for Microbial Growth Law''}
    }
\end{center}
\vspace{10pt}
\begin{table*}[bh]
    \centering
    \caption{{Notations in the main text}}
    \label{table:Symbols}
    \begin{tabular}{c|c}
    Symbol & Description  \\ \hline 
    $\mu$ & Growth rate {($=\hat{v}_\BM$)} \\
    $[S]$ &  Environmental concentration of (growth-limiting) nutrient $S$ \\
    $\Metabo, \Exchange$ & Set of metabolites / exchangeable metabolites ($\Exchange\subset \Metabo$) \\ 
    $\Const$ & Set of non-nutrient resources \\
    $\Reac$ & Set of reactions \\
    $\BM$ & Growth or biomass synthesis reaction ($\BM\in\Reac$) \\
    $\mathsf{S}, \mathsf{C}$ & Stoichiometry matrix ($\mathsf{S}:=\{\mathsf{S}_{m i}\}_{m\in\Metabo, i\in\Reac}$) and resource allocation matrix
    ($\mathsf{C}:=\{\mathsf{C}_{a i}\}_{a\in\Const, i\in\Reac}$) \\
    $I_a$ & Maximal intake of exchangeable metabolite $a\,({\in}\,\Exchange$) or total capacity for non-nutrient resource $a\,({\in}\,\Const$) \\ 
    ${\Income}$ & Availability of the resources (${\Income}:=\{I_a\}_{a\in\Exchange\cup\Const}$) \\ 
    $\tilde{\Income}$ & Availability of the resources other than nutrient $S$ ($\tilde{\Income}:=\{I_a\}_{a\in\Exchange\cup\Const\backslash\{S\}}$) \\ 
    $v_i$ & Non-negative flux of reaction $i\,(\in\Reac)$ \\ 
    $\vflux,\hat{\vflux}$ & Reaction fluxes ($\vflux\,{:=}\,\{v_i\}_{i\in\Reac}$) and its optimized solution ($\hat{\vflux}\,{:=}\,\{\hat{v}_i\}_{i\in\Reac}$)\\ 
    $\hat{y}_m$ & Shadow price of metabolite $m\,({\in}\,\Metabo)$ or non-nutrient resource $m\,({\in}\,\Const)$  \\ 
    ${\bf y}, \hat{{\bf y}}$ & Variables of dual problem~\eqref{eq:dual} (${\bf y}\,{:=}\,\{y_a\}_{a\in\Metabo\cup\Const}$) and its optimized solution (i.e., shadow prices $\hat{{\bf y}}\,{:=}\,\{y_a\}_{a\in\Metabo\cup\Const}$)
    \end{tabular}
\end{table*}
\vspace{20pt}
\twocolumngrid

\section{{Derivation of the dual problem~\eqref{eq:dual}}} \label{sec:dual}
To derive the dual problem, let us introduce matrices 
\begin{eqnarray*} 
\Nint&:=&-\mathsf{S}^{\Metabo\backslash\Exchange}
    = \{ -\mathsf{S}_{a i} \}_{a\in\Metabo\backslash\Exchange,i\in\Reac}, \\
\Nex&:=&\begin{pmatrix} -\mathsf{S}^{\Exchange} \\ \mathsf{C} \end{pmatrix}, 
\end{eqnarray*}
with $\mathsf{S}^{\Exchange} := \{ \mathsf{S}_{a i}  \}_{a\in\Exchange,i\in\Reac}$, 
and rewrite the linear programming (LP) problem~(\ref{eq:Balanced}-\ref{eq:ResourceAllocation}) as 
\begin{eqnarray*} \label{eq:CBM_formal}
    \underset{\vflux}{{\rm maximize}} \;v_\BM \;\; \mathrm{s.t.} \;\;
    \Nint\vflux={\bf 0}, \;\; \Nex\vflux\leq \Income , \;\; \vflux\geq{\bf 0}
    .
\end{eqnarray*}
Then, the dual problem to this primal LP program~(\ref{eq:Balanced}-\ref{eq:ResourceAllocation}) is formally derived as 
\begin{eqnarray*} \label{eq:dual2}
    \underset{{\bf y}\in\mathbb{R}^{\Metabo\cup\Const}}{\min}
    \begin{pmatrix}
        {\bf 0} \\
        \Income
    \end{pmatrix}
    \cdot {\bf y} \quad 
    \mathrm{s.t.} \quad
    \begin{pmatrix}
        {\Nint}^\top & {\Nex}^\top \\
        {\bf 0} & \mathsf{I}
    \end{pmatrix} {\bf y} \geq {\bf 1}_\BM,
\end{eqnarray*}
where $\mathsf{I}$ is the $(\Exchange\cup\Const)\,{\times}\,(\Exchange\cup\Const)$ identity matrix and ${\bf 1}_\BM\in\{0,1\}^{\Reac\cup\Exchange\cup\Const}$ denotes a vector where the $\BM$-th element is unity and all other elements are zero~\cite{warren2007duality,vanderbei1998linear,reznik2013flux}. This can be rewritten as 
\begin{align*}
    & \underset{{\bf y}\in\mathbb{R}^{\Metabo\cup\Const}}{{\rm minimize}}
    \sum_{m\in\Exchange\cup\Const}I_my_m \\
    &\quad \mathrm{s.t.} \quad
    - \sum_{m\in\Metabo} \mathsf{S}_{mi} y_m + \sum_{m\in\Const}\mathsf{C}_{mi} y_m \geq 0 \quad(i\in\Reac\backslash\{\BM\}),\\
    &\quad \quad \quad \;
    -\sum_{m\in\Metabo} \mathsf{S}_{m,\BM} y_m + \sum_{a\in\Const}\mathsf{C}_{m,\BM} y_m \geq 1,\\
    &\quad \quad \quad \;\;\; y_a\geq 0\;\;(a\in\Exchange\cup\Const).
\end{align*}
It is equivalent to the dual problem~\eqref{eq:dual} in the main text.

\subsection{A simple example of the duality}\label{sec:SimpleEx}
\begin{figure}[tb]
    \centering \includegraphics[width = 0.99\linewidth]{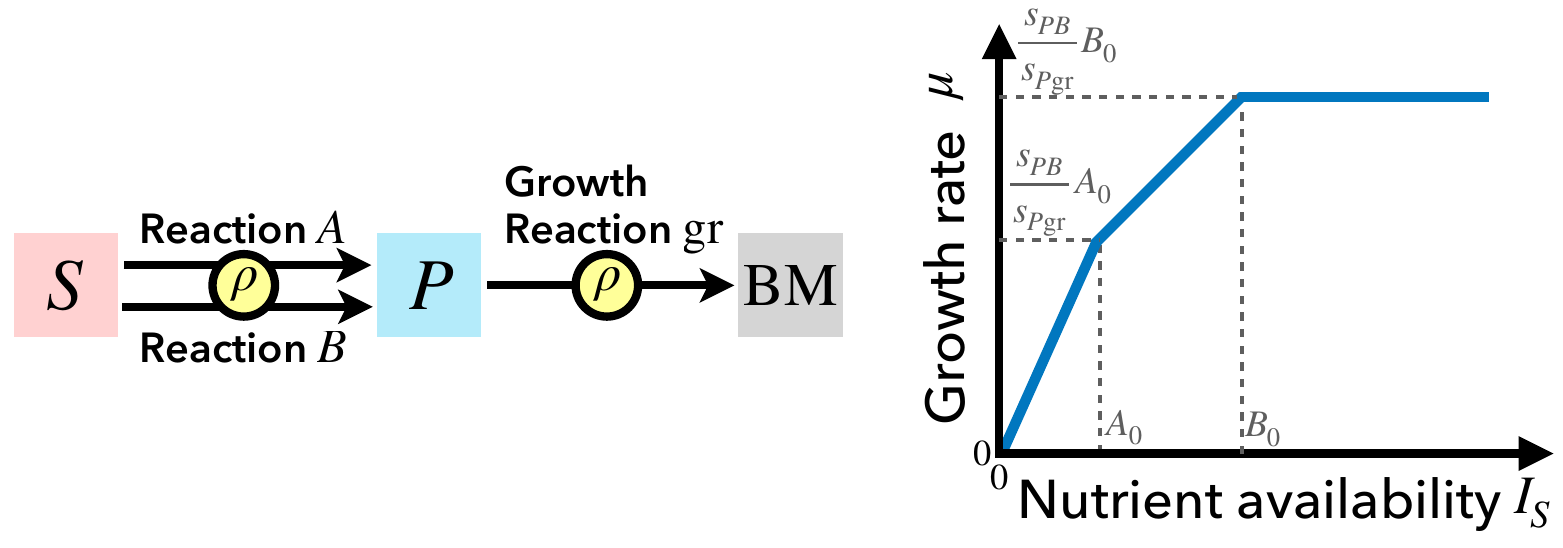}
    \caption{
        A simple example of metabolic systems. 
    }\label{fig:SI_toy_model}
\end{figure}
As a concrete example, we present here an analytically-solvable metabolic model (Fig.~\ref{fig:SI_toy_model}). 
It is a simple metabolic model where multiple pathways produce a common metabolite from a single nutrient: e.g., the co-utilization of respiration and fermentation~\cite{yamagishi2021microeconomics} or that of Embden–Meyerhoff–Parnass (EMP) and Entner–Doudoroff (ED) glycolytic pathways~\cite{EMP/ED}. 

This model consists of $2$ metabolites, $\Metabo=\{S,P\}$, and $1$ constraint $\Const=\{\rho\}$, as well as $3$ reactions, $\Reac=\{A,B,\BM\}$. 
Note that $\rho$ can be any kind of limited (non-nutrient) resource, such as the intracellular volume or solvent capacity~\cite{vazquez2010catabolic,OMbook}, the total amount of proteins~\cite{Hwa_OM}, and the total area of membrane surfaces~\cite{memRealEstate2}. 
We assume that $S$ is the only exchangeable metabolite, i.e., $\Exchange=\{S\}$. Reactions $A$ and $B$ both produce $P$ from $S$ by using non-nutrient resource $\rho$, and the growth reaction $\BM$ requires both $P$ and $\rho$ (Fig.~\ref{fig:SI_toy_model}). 
The stoichiometry matrix, $\mathsf{S}\in\mathbb{R}^{\Metabo\times\Reac}$, and the resource constraint matrix, $\mathsf{C}\in\mathbb{R}^{\Const\times\Reac}$, are then given as 
\begin{eqnarray*}
  \mathsf{S} &=& \begin{pmatrix}
    -1 & -1 & 0 \\
    s_{PA} & s_{PB} & -s_{P \BM}
    \end{pmatrix},
\\
  \mathsf{C} &=& \begin{pmatrix}
    \mathsf{C}_{\rho A} & \mathsf{C}_{\rho B} &  \mathsf{C}_{\rho \BM}
    \end{pmatrix}.
\end{eqnarray*}
As discussed in ref.~\cite{yamagishi2021microeconomics}, a trade-off between reactions $A$ and $B$ leads to the co-utilization of these pathways and switching between them. Therefore, we here assume a trade-off $s_{PA} > s_{PB}$ and $\mathsf{C}_{\rho A} > \mathsf{C}_{\rho B}$: i.e., reaction $A$ is more efficient in producing $P$ but requires more amount of constraint $\rho$ than the alternative reaction $B$. When this trade-off matters, the optimal flux $\hat{v}_A$ of reaction $A$ is suppressed by increasing maximal nutrient influx $I_S$. 

With matrices $\mathsf{S}$ and $\mathsf{C}$, the LP problem for CBM is represented as 
\begin{align} \label{eq:ToyModel_LP}
    \underset{\vflux\geq \textbf{0}}{\rm maximize}
    \;\; v_\BM
    \;\;\;
    & \mathrm{s.t.} 
    \\
    &\;\; -v_A -v_B + I_S \geq 0, \nonumber \\
    &\;\; s_{P A}v_A + s_{P B}v_B - s_{P \BM}v_\BM = 0, \nonumber \\
    &\;\; \mathsf{C}_{\rho A}v_A + \mathsf{C}_{\rho B}v_B + \mathsf{C}_{\rho \BM}v_\BM \leq I_\rho. \nonumber 
\end{align}
Its solution, or the optimal fluxes $\hat{\vflux}$, is calculated as 
\begin{align*}
    & \left(\hat{v}_{A},\hat{v}_{B}\right) \\
    & =\begin{cases} 
    \left( I_S, 0 \right)
    \quad & \mathrm{if}\;I_S\leq A_0\\
    \left(
        \frac{B_0-I_S}{B_0 - A_0}A_0
        ,
        \frac{I_S - A_0}{B_0 - A_0}B_0
    \right)
    \quad & \mathrm{if}\; B_0\geq I_S\geq A_0\\
    \left(0, B_0\right)
    \quad & \mathrm{if}\;I_S\geq B_0
    \end{cases},
\end{align*}
where $ A_0 := I_\rho / (s_{PA} + \mathsf{C}_{\rho A})$ and $ B_0 := I_\rho / (s_{PB} + \mathsf{C}_{\rho B})$. 
The dependence of the growth rate on $I_S$ is then calculated as 
\begin{align*}
    & \mu(I_S; I_\rho) =  \\
    & \begin{cases} 
        s_{PA}I_S
        & \mathrm{if}\;I_S\leq A_0\\
        s_{PA}\frac{B_0-I_S}{B_0 - A_0}A_0
        + s_{PB}\frac{I_S - A_0}{B_0 - A_0}B_0
        \quad & \mathrm{if}\; B_0\geq I_S\geq A_0\\
        s_{PB}B_0
        & \mathrm{if}\;I_S\geq B_0
    \end{cases}.
\end{align*}
It indeed satisfies monotonicity and concavity (see also Fig.~\ref{fig:SI_toy_model}). \\

The dual problem to the primal LP problem~\eqref{eq:ToyModel_LP} is 
\begin{align} \label{eq:ToyModel_dual}
    & \underset{\bf{y}\geq \textbf{0} }{\rm minimize}\quad 
    I_S y_S + I_{\rho} y_{\rho}  \\
    &\quad \mathrm{s.t.} \quad
    y_S - s_{P A} y_P 
     \geq 0, \nonumber \\
    &\quad\quad\quad\; 
    y_S - s_{P B} y_P 
    + \mathsf{C}_{\rho B} y_\rho \geq 0, \nonumber \\
    &\quad\quad\quad\;
    s_{P \BM} y_P 
    + \mathsf{C}_{\rho,\BM} y_\rho \geq 1. \nonumber 
\end{align}
{
    The dual problem is formal in nature and, unlike the original problem, is difficult to interpret biologically. However, an intuitive (economic) interpretation of this minimization problem is to find the smallest ``selling price that avoids incurring a loss,'' which is the solution of the dual problem or shadow price $\hat{{\bf y}}$. 
} 

The shadow price $\hat{y}_{S}$ of nutrient $S$ is calculated as 
\begin{align*} 
    & \hat{y}_{S}(I_S;I_\rho) = 
    \begin{cases} 
        s_{PA}
        \quad \mathrm{if}\;I_S\leq A_0\\
        \frac{ s_{PB}B_0 - s_{PA}A_0 }{B_0 - A_0}
        \quad \mathrm{if}\; B_0\geq I_S\geq A_0\\
        0
        \quad \mathrm{if}\;I_S\geq B_0
    \end{cases}.
\end{align*}
It is indeed monotonically decreasing. 

Note that the dual LP problem~\eqref{eq:ToyModel_dual} is different from minimizing the nutrient influx $I_S$ to achieve a given growth rate $\mu$, which is often considered in constraint-based modeling (CBM):
\begin{align*} 
    \underset{I_S,v_A,v_B\geq 0}{\rm minimize} \;\; I_S
    \quad
    \mathrm{s.t.}
    \;\;\;
    & -v_A -v_B + I_S \geq 0,\\
    & s_{P A}v_A + s_{P B}v_B - s_{P \BM}v_\BM = 0, \\
    & \mathsf{C}_{\rho A}v_A + \mathsf{C}_{\rho B}v_B + \mathsf{C}_{\rho \BM}v_\BM \leq I_\rho, \\
    & v_\BM \geq \mu,
\end{align*}
where the variables are nutrient influx $I_S$ and reaction fluxes $v_A$ and $v_B$, while growth rate $\mu$ and $I_\rho$ are given as parameters.

\section{Proof of the concavity of optimal objective function} \label{sec:concavity}
The concavity of optimal objective function $ \hat{v}_\BM(\Income) $ as a multivariable function of $\Income$
can be analytically proven as follows. 

From the LP duality theorem~\cite{vanderbei1998linear} and the dual problem~\eqref{eq:dual}, 
$$
    \hat{v}_\BM(\Income) =\underset{{\bf y}\in\mathbb{R}^{\Metabo\cup\Const}}{\min}\{ \Income \cdot \yec | \begin{pmatrix}
            -{\mathsf{S}}^\top & {\mathsf{C}}^\top
        \end{pmatrix} {\bf y} \geq {\bf 1}_\BM, \yec\geq {\bf 0} \}, 
$$ 
where $\yec$ is defined as $\yec:=\{y_a\}_{a\in\Exchange\cup\Const}$ and ${\bf 1}_\BM \in \{0,1\}^{\Reac}$ denotes a vector where the $\BM$-th element is unity and all other elements are zero. In contrast to the primal problem~(\ref{eq:Balanced}-\ref{eq:ResourceAllocation}), it is useful that the feasible solution space of the dual problem~\eqref{eq:dual} does not change when $\Income$ is varied. 

Thus, for arbitrary $0<t<1$ and $\Income_1,\Income_2$, 
\begin{align*} 
    & \hat{v}_\BM(t\Income_1+(1-t)\Income_2) \\
    & =  
    \underset{{\bf y}}{\min}\{ (t\Income_1+(1-t)\Income_2) \cdot \yec | \begin{pmatrix}
            -{\mathsf{S}}^\top & {\mathsf{C}}^\top
        \end{pmatrix} {\bf y} \geq {\bf 1}_\BM, \yec\geq {\bf 0} \} \\
    & \geq \underset{{\bf y}}{\min}\{ t\Income_1 \cdot \yec | \begin{pmatrix}
            -{\mathsf{S}}^\top & {\mathsf{C}}^\top
        \end{pmatrix} {\bf y} \geq {\bf 1}_\BM, \yec\geq {\bf 0} \} \\
    & \quad + \underset{{\bf y}}{\min}\{ (1-t)\Income_2 \cdot \yec | \begin{pmatrix}
            -{\mathsf{S}}^\top & {\mathsf{C}}^\top
        \end{pmatrix} {\bf y} \geq {\bf 1}_\BM, \yec\geq {\bf 0} \} \\
    & = t\hat{v}_\BM(\Income_1) + (1-t)\hat{v}_\BM(\Income_2). 
\end{align*}
This is the definition of the concavity.

\subsection{Proof of the concavity in nonlinear CBM}\label{sec:ConicFBA}
The above proof of the convexity in linear programs using the duality theorem can be extended to nonlinear convex programs.

In general, by adding $M$ convex constraints to the LP program~(\ref{eq:Balanced}-\ref{eq:ResourceAllocation}) in CBM, convex optimization problems in CBM can be formulated as:
\begin{eqnarray}
    \underset{\vflux,{\bf x}\geq{\bf 0}}{\rm maximize} \;\; v_\BM \quad \mathrm{s.t.} 
    && 
        \Nint\vflux={\bf 0},  \nonumber \\ 
    && 
        \Nex\vflux\leq \Income ,  \label{eq:nonlinear_CBM_formal}\\ 
    && 
        g_\alpha(\vflux,{\bf x}) \leq 0\;\; (\alpha=1,\cdots,M),  \nonumber
\end{eqnarray}
Here, the variables ${\bf x}$ typically represent metabolite concentrations, but can include other variables; $g_\alpha(\vflux,{\bf x})$ is assumed to be a convex function and represents, for example, constraints in the relationship between concentrations ${\bf x}$ and fluxes $\vflux$.

An example of a nonlinear convex optimization problem in CBM is Conic FBA~\cite{taylor2024convex}, which incorporates the relationship between reaction fluxes and metabolite concentrations as constraints and models cellular metabolism and growth as a second-order cone program. 
In this method, the Michaelis–Menten kinetics is represented as a convex function $g_\alpha(v_\alpha,{\bf x}) $ (see refs.~\cite{taylor2024convex,taylor2021second} for more details). \\

To prove the convexity of $\hat{v}_\BM(\Income)$, it is convenient to consider the Lagrange dual problem of the convex program~\eqref{eq:nonlinear_CBM_formal}:
\begin{align} \label{eq:nonlinear_CBM_dual}
    & \underset{{\bf y}}{\rm minimize} \;\;
    \sum_{a\in\Exchange\cup\Const} I_a y_a 
     + \underset{\vflux,{\bf x}\geq{\bf 0}}{\max}\sum_\alpha g_\alpha(\vflux,{\bf x})y_\alpha\\
    & \mathrm{s.t.} \;\;
    \begin{pmatrix}
        {\Nint}^\top & {\Nex}^\top \\
        {\bf 0} & \mathsf{I}
    \end{pmatrix} {\bf y} \geq {\bf 1}_\BM, 
    \;\; y_\alpha \geq 0 \;\;(\alpha=1,\cdots,M). 
    \label{eq:nonlinear_CBM_dual_const}
\end{align}
As with the LP dual problem, note that the parameter $\Income$ appears only as a coefficient in the objective function and does not affect the feasible solution. 
From the strong duality theorem~\cite{luenberger1984linear}, for arbitrary $0<t<1$ and $\Income_1,\Income_2$, 
\begin{align*} 
    & \hat{v}_\BM(t\Income_1+(1-t)\Income_2) \\
    & =  
    \underset{{\bf y}}{\min}\{ (t\Income_1+(1-t)\Income_2) \cdot \yec + \sum_\alpha g_\alpha(\hat{\vflux},\hat{\bf x})y_\alpha | \eqref{eq:nonlinear_CBM_dual_const} \} \\
    & \geq \underset{{\bf y}}{\min}\{ t\Income_1 \cdot \yec + t \sum_\alpha g_\alpha(\hat{\vflux},\hat{\bf x})y_\alpha | \eqref{eq:nonlinear_CBM_dual_const} \} \\
    & \quad + \underset{{\bf y}}{\min}\{ (1-t)\Income_2 \cdot \yec + (1-t) \sum_\alpha g_\alpha(\hat{\vflux},\hat{\bf x})y_\alpha | \eqref{eq:nonlinear_CBM_dual_const} \} \\
    & = t\hat{v}_\BM(\Income_1) + (1-t)\hat{v}_\BM(\Income_2). 
\end{align*}
This is the definition of the concavity. 

The monotonic increase in the growth kinetics curve $\mu(I_S;\tilde{\Income})$ is also evident. 

\begin{figure}[t]
    \centering 
    \includegraphics[width=0.9\linewidth, clip]{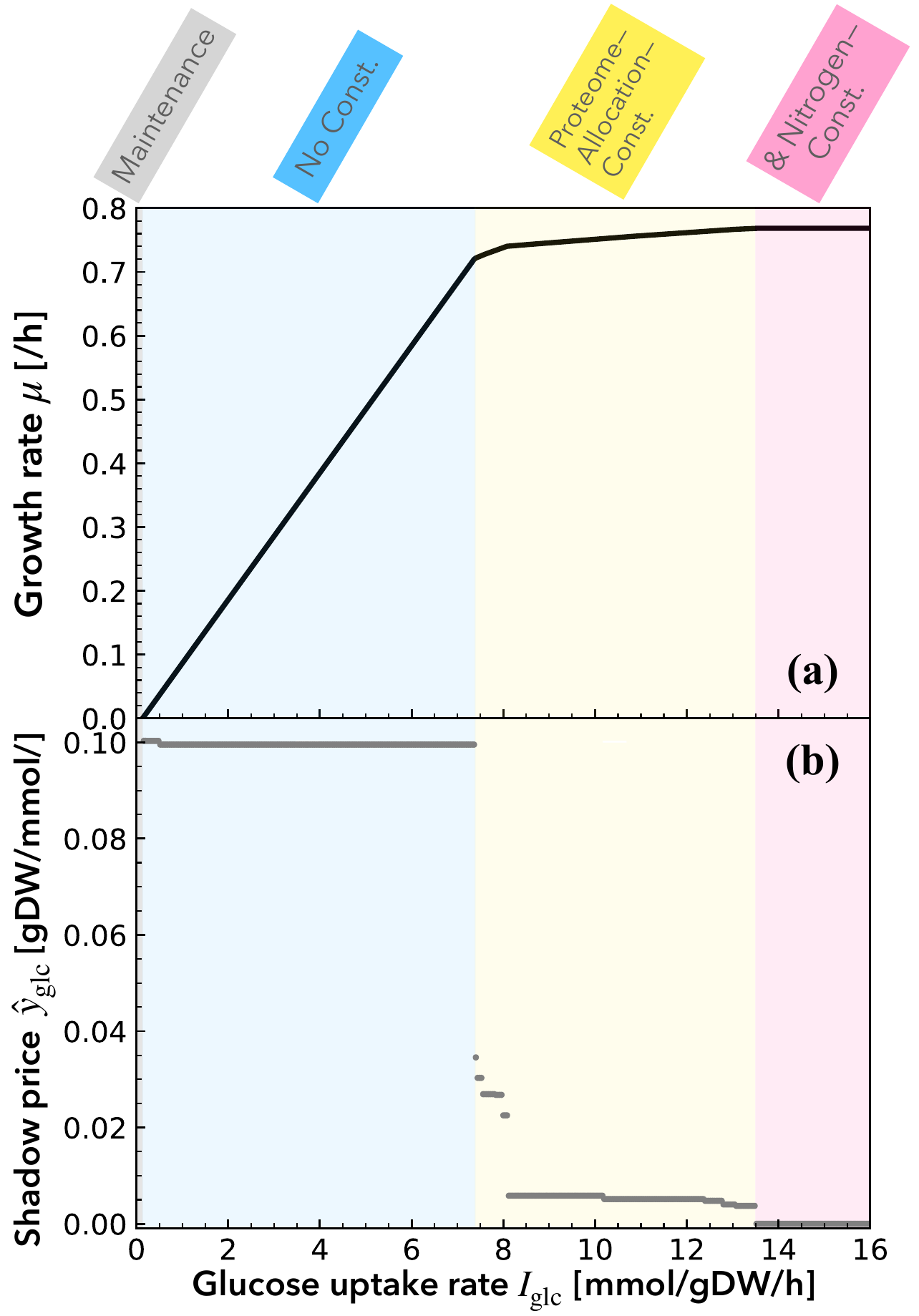}
    \caption{
    (a) Growth rate $\mu$ and (b) shadow price $\hat{y}_{\rm glc}$ of glucose as a function of carbon source availability $I_\mathrm{glc}$ with $I_{\rm amm}\,{=}\,8.3$. Numerical calculations of a CBM method {with proteome allocation}, constrained allocation flux balance analysis (CAFBA)~\cite{mori2016constrained}, were performed using the genome-scale \textit{E. coli} iJO1366 model~\cite{orth2011comprehensive} and the COBRApy package~\cite{ebrahim2013cobrapy}. {See also Appendix~\ref{sec:CBMs}.}
    }\label{fig:general_ex}
\end{figure}

\section{Details of numerical simulations}\label{sec:CBMs}
We used several CBM methods for numerical demonstration of our general proof (Figs.~\ref{fig:VariousFBAs}~and~\ref{fig:general_ex2} in the main text). 
Formally, all of the below methods can be formulated as Eqs.~(\ref{eq:Balanced}-\ref{eq:ResourceAllocation}) in the main text, we here explain more details of each method. 

The mass balance due to intracellular stoichiometry is identical in all the methods, while the additional non-nutrient resources $\Const$ and corresponding constraints are different across methods, as explained below. 

\subsection{Constrained Allocation Flux Balance Analysis (CAFBA)~\cite{mori2016constrained}}\label{sec:CAFBA}
In the method of CAFBA~\cite{mori2016constrained}, the allocation of the proteins associated with ribosome, biosynthetic enzymes, carbon transport, and housekeeping reaction processes is introduced as $\phi_R$, $\phi_E$, $\phi_C$, and $\phi_Q$, respectively. 
The sum of these proteome fractions is normalized as $\phi_R + \phi_E + \phi_C + \phi_Q = 1$. 

Each proteome fraction $\phi_R$, $\phi_E$, or $\phi_C$ (except for $\phi_Q$ for housekeeping processes) varies with environmental conditions and physiological states as follows: the ribosome sector $\phi_R$ increases proportionally to biomass synthesis rate $v_\BM$ as $\Delta\phi_R = \mathsf{C}_{\phi, \BM}v_\BM$; the biosynthetic enzyme sector $\phi_E$ changes proportionally to the reaction flux $v_i$ such that $\Delta\phi_E = \sum_i \mathsf{C}_{\phi, i} v_i$; the carbon transport sector $\phi_C$ changes proportionally to the carbon intake flux as $\Delta\phi_C = \mathsf{C}_{\phi, C}v_C$. 
Accordingly, the constraint in CAFBA for partitioning the flexible proteome fractions is represented as: 
\begin{eqnarray*}\label{eq:CAFBA}
    \Delta\phi_R + \Delta\phi_E + \Delta\phi_C = \sum_{i \in \Reac} \mathsf{C}_{\phi, i} v_i \leq I_\phi. 
\end{eqnarray*}

$I_\phi$ is set to $0.4$ and the parameter values of $\mathsf{C}_{\phi, i}$ are calibrated according to the methods in ref.~\cite{takano2023architecture}.

\subsection{Flux Balance Analysis with Molecular Crowding (FBAwMC)~\cite{beg2007intracellular,vazquez2008impact}}\label{sec:Vazquez}
The method of FBAwMC~\cite{beg2007intracellular,vazquez2008impact} incorporates a limited solvent capacity for the allocation of metabolic enzymes. 
Given that the enzyme molecules have a finite molar volume $V_i$, only a finite number of them fit in a given cell volume $V$. That is, if $n_i$ is the number of moles of enzyme $i$, 
$$
    \sum_{i \in \Reac} V_i n_i \leq V.
$$ 
It is assumed that an enzyme concentration $E_i:= n_i/M$ (moles per unit mass; $M$ is cell mass) results in a flux $v_i = b_iE_i$ for each reaction $i$, where the parameter $b_i$ is determined by the reaction mechanism, kinetic parameters, and metabolite concentrations. 
Then, the enzyme concentration constraint due to molecular crowding is reflected as the following metabolic flux constraint:
\begin{eqnarray}\label{eq:FBAwMC}
    \sum_{i \in \Reac} \mathsf{C}_{V,i} v_i\leq 1 =: I_V ,
\end{eqnarray}
where $\mathsf{C}_{V,i} := \frac{MV_i}{Vb_i}$. 

According to the original paper~\cite{beg2007intracellular}, $\mathsf{C}_{V,i}$'s are given as random values from the gamma distribution $P(a)$ with the average $\langle a \rangle \simeq 0.0031$ [h·g/mmol] and the shape parameter $\beta=3$; note that for $\beta\gg1$, we obtain a distribution that is almost concentrated around $a=\langle a \rangle$. 
In the numerical simulation of Fig.~2 in the main text, The stoichiometry matrix $\mathsf{S}$ is based on the genome-scale \textit{E. coli} iJO1366 model~\cite{orth2011comprehensive}. 

\subsection{Specific Membrane Surface Area-constrained Flux Balance Analysis (sMSAc-FBA)~\cite{carlson2024cell}}\label{sec:MemEstate} 
The method of sMSAc-FBA~\cite{carlson2024cell} incorporates the limited capacity of specific membrane surface area (sMSA) to host proteins. 
Given the finite surface area-to-volume ratio (SA:V)~\cite{memRealEstate2, carlson2024cell}, denoted as $I_{\rm MRE}$, the corresponding constraint is expressed as:
\begin{eqnarray}\label{eq:sMSAc-FBA}
    \sum_{i\in\Reac} \mathsf{C}_{{\rm MRE}, i} v_i \leq I_{\rm MRE}, 
\end{eqnarray}
where ``membrane real estate'' coefficient $\mathsf{C}_{{\rm MRE}, i}$ ${\rm [h\cdot nm^2/mmol]}$ represents the membrane area used by active enzyme for reaction $i$. 

According to ref.~\cite{carlson2024cell}, the parameter values of $\mathsf{C}_{{\rm MRE}, i}$ were chosen with the intracellular stoichiometry derived from the \textit{E. coli} core model, and $I_{\rm MRE}\simeq 2.8$ ${\rm [nm^2/gDW]}$ is specified.

\end{document}